\documentclass[useAMS,usenatbib]{mn2e}
\usepackage{amssymb,latexsym,graphicx,natbib,times}
\newcommand{\mi}{\relax \ifmmode {\mu{\mbox m}}\else $\mu$m\fi}
\newcommand{\hii}{\relax \ifmmode {\mbox H\,{\scshape ii}}\else H\,{\scshape ii}\fi}
\newcommand{\sii}{\relax \ifmmode {\mbox S\,{\scshape ii}}\else S\,{\scshape ii}\fi}
\newcommand{\nii}{\relax \ifmmode {\mbox N\,{\scshape ii}}\else N\,{\scshape ii}\fi}
\newcommand{\neii}{\relax \ifmmode {\mbox Ne\,{\scshape ii}}\else Ne\,{\scshape ii}\fi}
\newcommand{\neiii}{\relax \ifmmode {\mbox Ne\,{\scshape iii}}\else Ne\,{\scshape iii}\fi}
\newcommand{\oiii}{\relax \ifmmode {\mbox O\,{\scshape iii}}\else O\,{\scshape iii}\fi}
\newcommand{\oii}{\relax \ifmmode {\mbox O\,{\scshape ii}}\else O\,{\scshape ii}\fi}
\newcommand{\oi}{\relax \ifmmode {\mbox O\,{\scshape i}}\else O\,{\scshape i}\fi}
\newcommand{\ha}{\relax \ifmmode {\mbox H}\alpha\else H$\alpha$\fi}

\newcommand{\hep}{\relax \ifmmode {\mbox H}\epsilon\else H$\epsilon$\fi}
\newcommand{\hdel}{\relax \ifmmode {\mbox H}\delta\else H$\delta$\fi}
\newcommand{\hgam}{\relax \ifmmode {\mbox H}\gamma\else H$\gamma$\fi}

\newcommand{\pa}{\relax \ifmmode {\mbox Pa}\alpha\else Pa$\alpha$\fi}
\newcommand{\hb}{\relax \ifmmode {\mbox H}\beta\else H$\beta$\fi}
\newcommand{\rdostres}{\relax \ifmmode {\,\mbox{R}}_{\rm 23}\else \,\mbox{R}$_{\rm 23}$\fi}
\newcommand{\ergs}{\relax \ifmmode {\,\mbox{erg\,s}}^{-1}\else \,\mbox{erg\,s}$^{-1}$\fi}
\newcommand{\me}{\relax \ifmmode {\,}^{-1}\else \,$^{-1}$\fi}
\newcommand{\degree}{\hbox{$^\circ$}}

\newcommand{\kms}{km\,s$^{-1}$}

\newcommand{\msun}{\relax \ifmmode {\,\mbox{M}}_{\odot}\else \,\mbox{M}$_{\odot}$\fi}
\newcommand{\zsun}{Z$_{\odot}$}
\newcommand{\cmtres}{\relax \ifmmode {\,\mbox{cm}}^{-3}\else \,\mbox{cm}$^{-3}$\fi}
\newcommand{\cmdos}{\relax \ifmmode {\,\mbox{cm}}^{-2}\else \,\mbox{cm}$^{-2}$\fi}
\newcommand{\cmseis}{\relax \ifmmode {\,\mbox{cm}}^{-6}\else \,\mbox{cm}$^{-6}$\fi}
\newcommand{\hi}{\relax \ifmmode {\mbox H\,{\scshape i}}\else H\,{\scshape i}\fi}

\title[Physical properties of NGC~595]{Spatially resolved study of the physical properties of the ionized
 gas in NGC~595 }
 
\author[M. Rela\~no, A. Monreal-Ibero, J. M. V\' ilchez and R. C. Kennicutt]{M. Rela\~no$^{1}$\thanks{E-mail:
mrelano@ast.cam.ac.uk (MR); amonreal@eso.org (AMI), jvm@iaa.es (JMV) and robk@ast.cam.ac.uk}, 
A. Monreal-Ibero$^{2}$, J. M. V\' ilchez$^{3}$ and R. C. Kennicutt$^{1}$ \\
$^{1}$Institute of Astronomy, University of Cambridge, Madingley Road, Cambridge, CB3 0HA, UK\\
$^{2}$European Organisation for Astronomical Research in the
Southern Hemisphere, Karl-Schwarzschild-Strasse 2, D-85748 Garching bei M\"unchen, Germany \\
$^{3}$Instituto de Astrof\'isica de Andaluc\' ia, CSIC, C/ Camino Bajo de Hu\' etor, 50, 18008, Granada, Spain} 

\begin{document}

\date{}

\pagerange{\pageref{firstpage}--\pageref{lastpage}} \pubyear{2002}

\maketitle

\label{firstpage}

\begin{abstract}
We present Integral Field Spectroscopy (IFS) of NGC~595, one of the most luminous \hii\ regions in M33. This type of observations allows us to study the variation of the principal emission-line ratios across the surface of the nebula. At each position of the field of view, we fit the main emission-line features of the spectrum within the spectral range 3650-6990~\AA, and create maps of the principal emission-line ratios for the total surface of the region. The extinction map derived from the Balmer decrement and the {\it absorbed} \ha\ luminosity show good spatial correlation with the 24~\mi\ emission from {\it Spitzer}. We also show here the capability of the IFS to study the existence of Wolf-Rayet (WR) stars, identifying the previously catalogued WR stars and detecting a new candidate towards the north of the region. The ionization structure of the region nicely follows the \ha\ shell morphology and is clearly related to the location of the central ionizing stars. The electron density distribution does not show strong variations within the \hii\ region nor any trend with the \ha\ emission distribution. We study the behaviour within the \hii\ region of several classical emission-line ratios proposed as metallicity calibrators: while [\nii]/\ha\  and [\nii]/[\oiii] show important variations, the \rdostres\ index is substantially constant across the surface of the nebula, despite the strong variation of the ionization parameter as a function of the radial distance from the ionizing stars. These results show the reliability in using the R$_{\rm 23}$ index to characterize the metallicity of \hii\ regions even when only a fraction of the total area is covered by the observations.

\end{abstract}


\section{Introduction}
Emission-line spectra of Galactic and extragalactic HII regions are normally obtained to characterize their physical properties, to extract information of the stellar populations that ionize them \citep{Osterbrock:2006p498} as well as to study the variation of the physical properties of the star-forming regions across the galaxy disc (e.g. \citealt*{McCall:1985p496}; \citealt{Vilchez:1988p505}; \citealt{VilaCostas:1992p506}). 
The \hii\ regions are far from the idealized spherical ionized gas clouds with homogenous physical properties, as has clearly been shown in detailed studies of the nearest Galactic and extragalactic \hii\ regions: e. g. Orion Nebula (\citealt{Baldwin:1991p116}; \citealt{Kennicutt:2000p110}; \citealt{Sanchez:2007p501}), 30~Doradus (\citealt{Mathis:1985p115}; \citealt{Kennicutt:2000p110}) and NGC~604 (\citealt{GonzalezDelgado:2000p486}; \citealt{MaizApellaniz:2004p493}), among others. The derived physical properties, generally obtained from long-slit
spectroscopic measurements centred on the most intense knots, are not necessarily representative of the conditions in all the locations within the regions (\citealt{Oey:2000p124}; \citealt{Oey:2000p121}). The variations of the physical properties are especially seen in star-forming regions where the action of the massive stars strongly influence the surrounding medium (e.g. \citealt*{Kobulnicky:1996p489}, 1997). Moreover, in some cases the gas within the \hii\ regions is ionized by multiple star clusters which are not always centrally located. This will produce different temperature and ionization structures than the normally assumed configurations from spherical models (\citealt*{Ercolano:2007p482}; \citealt{Jamet:2008p570}). 

In order to study the variations of the physical properties within the \hii\ regions, the usual technique carried out up to now consists of a set of spectroscopic observations with slits located at different positions in the region and covering as much surface as possible. This technique has limitations however: it requires a large amount of observation time and does not cover, in general, the complete face of the region; thus interpolations of the observational parameters in the gap between slits need to be made. Integral Field Spectroscopy (IFS) overcomes these limitations and offers the opportunity of extracting spectroscopic information at different
positions across a continuous field of view. Therefore, this kind of surveys allows the study of the variation of the physical
properties within star-forming regions. 

We present here IFS observations of NGC~595, the second most luminous \hii\ region in M33. 
NGC~595 presents an angular size of $\sim$1 arcmin, which at the distance of M33 (840~kpc; \citealt*{Freedman:1991p485}) translates into a linear physical size of $\sim$200~pc. This makes the region particularly suitable for being studied with the IFS technique. Although the current IFS facilities usually have a relatively small field of view, the proximity of this region (which implies a relatively high surface brightness) makes possible to map it in a reasonable amount of observing time.

NGC~595 has an \ha\ shell morphology that shows the action of the stellar winds of the massive stars located in its interior. The stellar content studied by \citet*{Malumuth:1996p494} using optical photometry reveals the existence of $\sim$250 OB-type stars,  $\sim$13 supergiants and 10 candidate Wolf-Rayet (WR) stars. These authors derived an age of 4.5 $\pm$ 1.0~Myr for the stellar cluster, which is consistent with the age predicted from the synthesis of integrated spectra in the far-ultraviolet (FUV) wavelength range \citep{Pellerin:2006p499}. Recently, \citet{Drissen:2008p481} spectroscopically confirmed nine of the WR candidates previously identified by \citet*{Drissen:1993p105} 
using photometric observations and discarded two possible WR candidates from their sample. The physical properties for NGC~595 have been derived with long-slit spectroscopic observations located at the most intense knots within the region \citep{Vilchez:1988p505} and recently with echelle spectroscopy \citep{Esteban:2009p563}. \citet{Vilchez:1988p505} obtained a temperature of T$_{\rm e} \sim$ 8000~K, an electron density consistent with the low density limit and a metallicity of 12 + log[O/H] = 8.44 $\pm$ 0.09 (Z = 0.6\zsun\footnote{We assume solar abundance from \citet*{Asplund:2005p404}, 12 + log[O/H] = 8.66}), which was recently confirmed by \citet{Esteban:2009p563}. The extinction within the region has been studied previously by other authors: \citet*{Bosch:2002p478} obtained the extinction using the 
\ha/\hb\ emission-line ratio and \citet*{Viallefond:1983p504} derived the extinction with the \ha\ and thermal radio emission ratio. \citet{Relano:2009p558} revised the extinctions using radio-to-\ha\ emission and derived new values using the 24~\mi-to-\ha\ integrated emission ratio, finding consistent results from both methods. They also found differences in the emission distributions at several wavelength ranges: while the infrared emission at 
24~\mi\ and \ha\ are spatially correlated with each other, the FUV emission is located within the observed \ha\ shell structure of the region. They suggest that the dust emitting at 24~\mi\ will probably be mixed with the ionized gas of the region and will be heated by the central ionizing stars.   

In this paper, we analyse IFS observations covering the whole face of NGC~595. We are able to perform a detailed analysis of the principal emission-line ratios which are widely used to describe the main properties of \hii\ regions such as extinction and density structure, ionization parameter, metallicity, existence of shocks and evolutionary state. With these observations, we are in a position to test the influence of the geometry and the stellar distribution within the region on the integrated fluxes of the emission lines, as has been predicted by the models. 
The comparison of our results with those obtained from long-slit spectroscopy makes it possible to estimate the bias of the observations when only one part of the region, generally the most intense, is covered. Finally, the power of the IFS and the quality of the observations presented here allow us to analyse the WR stellar content of the total surface of the region and to make a complete census of this stellar population in NGC~595. 

\section{Observations and Data Reduction}
\subsection{Observations}
The data were obtained with the \emph{Potsdam Multi Aperture
Spectrophotometer}, PMAS \citep{Roth:2005p500} at the 3.5~m telescope in
Calar Alto on the night of 2007 October 10. We used the lens array (LARR), made of
16 $\times$ 16$^2$ elements, with the magnification scale of
1.0 $\times$ 1.0 arcsec$^2$. This configuration allows observation at 
high spatial resolution and with a filling factor of 1.00 \citep{Kelz:2006p487}.
We used the V300 grating, which provides with relatively low
spectral resolution (3.40~\AA~pix$^{-1}$) but allows to cover the
whole optical spectral range (3650--6990~\AA).  

\begin{figure}
\includegraphics[width=0.45\textwidth]{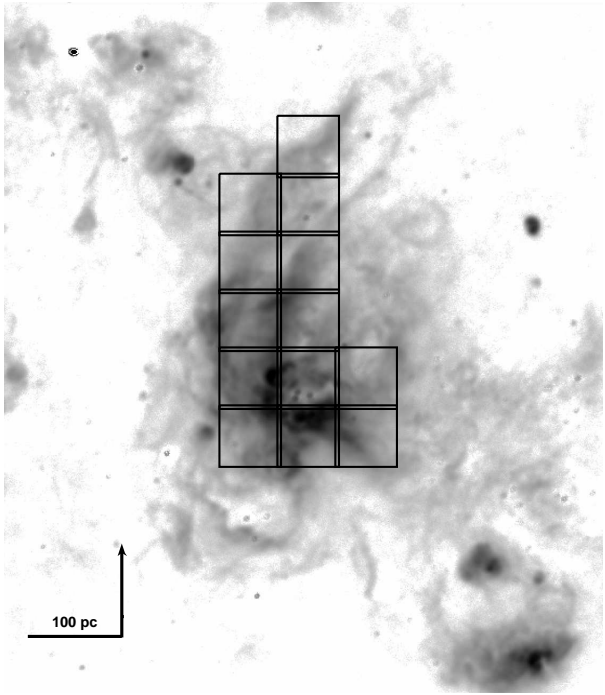}
\caption{Mosaic to map NGC~595 overplotted on a continuum-substracted 
  H$\alpha$ direct image from NOAO (National Optical Astronomy Observatory) Science Archive \citep{Massey:2007p492}. 
  The orientation is north up and east to the left. The H$\alpha$ image is shown in logarithmic stretch to better
  enhance all the morphological features of the H\,\textsc{ii} region.}
\label{apuntado}
\end{figure}

Given the relative small field of view of the LARR in comparison to
the angular size of NGC~595, we made a mosaic of 13 tiles  
to map the whole \hii\ region, covering a field of view of 47 $\times$ 92 arcsec$^2$, corresponding $\sim$174 $\times$ 340~parsec$^2$ at the distance of M33. The distribution of the different tiles is shown in Fig. \ref{apuntado} overplot on an
H$\alpha$ emission-line image from NOAO (National Optical Astronomy Observatory) Science Archive \citep{Massey:2007p492}. In order to have all
the data in common units, we made a 2.0 arcsec overlapping between contiguous
tiles.

We obtained 3-4 exposures of 400~s each per tile, depending on the
transparency of the sky and the surface brightness of the mapped region. Under
non-photometric conditions, seeing ranged typically between 1.2 and 1.8 arcsec, while the air
masses ranged between 1.9 and 1.0.
In addition to the science frames, continuum and HgNe arc lamps
exposures as well as 
sky background frames were obtained interleaved among the science
frames. This allowed us to minimize the effects due to flexures in the
instrument and ensure a proper sampling of background variations.

Finally, exposures of the  spectrophotometric standard-stars
BD+28D4211, HR153, G191-B2B were obtained in order to correct for the instrument
response and perform a relative flux calibration.

\subsection{Data reduction}

The data reduction follows the process described in \citet{AlonsoHerrero:2009p569} with
some minor modifications.  
We performed the basic reduction of each individual frame
using a set of homemade scripts created under the \emph{IRAF} 
enviroment\footnote{The Image Reduction and Analysis Facility
  \emph{IRAF} is distributed by the National Optical Astronomy
  Observatories which is 
  operated by the association of Universities for Research in
  Astronomy, Inc. under cooperative agreement with the National
  Science Foundation.}. The PMAS's CCD presents some structure, so bias was removed using a
masterbias made out the combination of 10 bias frames. We then
identified the location of the spectra in the CCD using continuum lamp
exposures (i.e. tracing) and extracted them.

We performed a wavelength calibration using the HgNe lamp
exposures. In order to 
estimate the quality of this calibration, we fitted a Gaussian profile to the
[O\,\textsc{i}]$\lambda$5577 skyline in each spectrum. The standard
deviation for the centroid of the line in each individual exposure
ranged typically between 0.1 and 0.4~\AA.  We then corrected from the
lens to lens sensitivity 
variations by creating response images from the continuum lamp
exposures.

At this stage, cosmic rays were removed in each individual
frame using the spectroscopic version of the L.~A. Cosmic routine
\citep{vanDokkum:2001p503}. Input parameters were chosen in order to eliminate as
most cosmic rays as possible. In the few possible cases where some spectral ranges close to emission lines were affected by cosmic rays (namely H$\alpha$ and, in some particular
cases, H$\beta$ and [O\,\textsc{iii}]$\lambda$5007), these were
removed manually.

The next step was the relative flux calibration performed
using the G191-B2B standard star exposure, which was observed at very low airmass (i.e. 1.06). The flux
calibration was done using the \emph{IRAF} tasks \texttt{standard, sensfunc and calibrate}. In order to estimate the uncertainties associated to this calibration, we made a comparison of the sensitivity functions derived for G191-B2B and for the other two standards. For the spectral range of $4500--7000$~\AA, uncertainties were $\sim$5 per cent, while in the bluer part of the spectra (i.e. $<$4500~\AA), they could reach 15 per cent. Note, however, that since it was not strictly necessary for the goals of the present work and the night was not photometric, we did not perform any absolute flux calibration.

For sky subtraction, we created a high signal-to-noise (S/N) spectrum
for each of the interleaved sky exposures by combining its 256
spectra. This collection of combined spectra sampled the sky variations along the night. 
The sky in each individual tile was subtracted using the sky spectrum obtained within $\sim$1 h of the time 
of the object exposure. Because of the non-photometric conditions, small
scaling factors (i.e. between 0.8 and 1.5) had to be included in the
sky spectra in order to minimize the skyline residuals.

Finally, we made a unique data cube for the whole H\,\textsc{ii} region
by mosaicking the individual frames using the offsets commanded at the
telescope. For this purpose, we used the \texttt{create\_mosaic.pro}
routine from the P3d package \citep{Becker:2006p510,Roth:2005p500} which corrects from
the effect of the differential atmospherical difraction by using the
expresion given in \citet{Filippenko:1982p483}. All the frames were put to common
units by using the overlapping bands between the individual tiles.

\subsection{Line fitting and map creation}
\label{line_fit}

We used MPFITEXPR algorithm\footnote{See http://purl.com/net/mpfit} implemented by C. B. Markwardt in IDL to fit the emission lines in each spatial element (hereafter, \emph{spaxel}) \citep{Markwardt:2009p495}. The advantage of this algorithm is the simplicity of imposing restrictions in the parameters of the fit, which is carried out using Levenberg-Marquardt least-squares minimization. 

For each \emph{spaxel} we selected the wavelength range that includes the emission lines we are interested in fitting and then performed the minimization of the observed profile to a Gaussian profile describing the emission line plus a one-degree polynomial function for the continuum subtraction. As an illustration, for the \ha+[\nii] emission lines we selected a wavelength range of 6500--6650~\AA, imposed a fixed wavelength separation between the \ha, [\nii]$\lambda$6548 and [\nii]$\lambda$6584 given by the redshift provided by NED\footnote{http://nedwww.ipac.caltech.edu} and performed the fit keeping the same linewidth for all the emission lines and a nitrogen line ratio of 3.  We estimated the noise of the corresponding spectrum using the standard deviation of the adjacent continuum. The noise defined in this way was used to obtain the S/N for each emission line fitted. We performed the fit in each \emph{spaxel} and rejected in the procedure those fits with a S/N  smaller than 5. The remaining spectra were visually inspected and bad fits were interactively rejected. 

The same procedure was applied to all the observed emission lines. The flux for each emission line was obtained by integrating the area defined by the best-fitting Gaussian profile. Then we used the derived flux and the position within the data cube for each  \emph{spaxel} to create an image (a \emph{map}) that can be treated with standard astronomical software.
We performed the astrometry of the emission-line maps using a continuum map close to the \ha\ emission line derived in the fitting procedure and a R-Band image obtained from the NOAO Science Archive \citep{Massey:2006p517} to identify the most intense stellar clusters in the continuum map. The accuracy of the astrometry is better than 1 arcsec, the angular size of the \emph{spaxel} in our field of view. 

\section{Integrated properties}
\label{secint}

\begin{figure*}
\begin{center}
\begin{minipage}[]{0.7\textwidth}
\includegraphics[width=\textwidth]{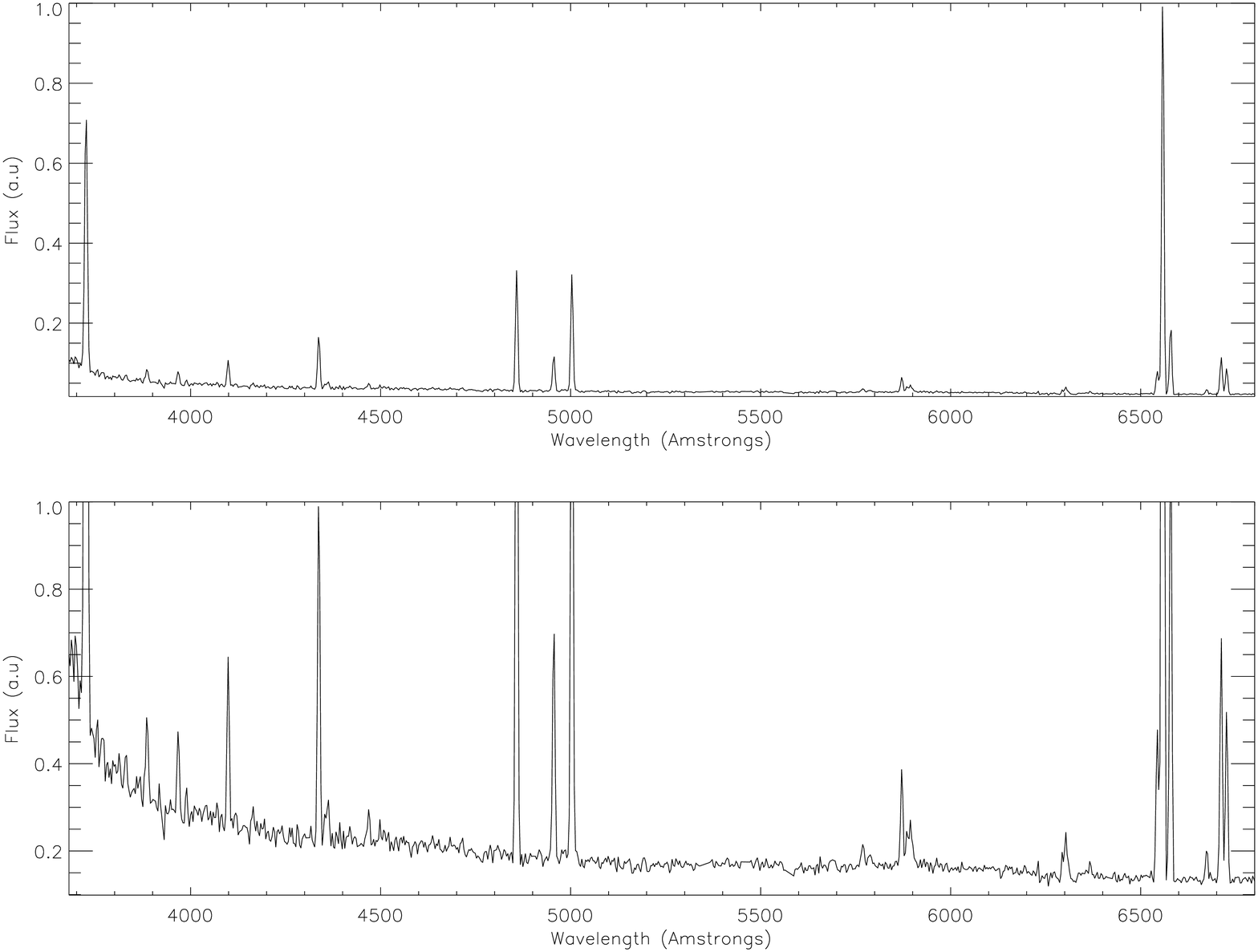}
\end{minipage}
\caption{Integrated spectrum of NGC~595 obtained by coadding the signal of all the \emph{spaxels} in the field of view with two different normalizations to better visualize the different observed emission lines. Fluxes are in arbitrary units, and the main detected emission lines are listed in Table~\ref{lambdaint}. Some residuals at $\sim$5590~\AA\ and $\sim$6300~\AA\  produced in the sky subtraction procedure under non-photometric conditions are seen in the integrated spectrum.}
\label{espec_integ}
\end{center}
\end{figure*}

Before discussing the emission-line maps of NGC~595, we first analyse the spectrum for the whole \hii\ region. In Fig.~\ref{espec_integ} we show the integrated spectrum in arbitrary units obtained by coadding the signal from the $\sim$3000 \emph{spaxels}. We use two different normalizations in the figure to show all the observed emission lines. We checked our method  (Section~\ref{line_fit}) by comparing the measured fluxes with those derived using the task \texttt{splot} from IRAF. We find differences in the fluxes of $\sim$1 per cent, which shows that our method is reliable and gives consistent results. We corrected the fluxes using the observed \ha/\hb, \hgam/\hb, \hdel/\hb, and \hep/\hb\ ratios and the mean extinction law of \citet{Savage:1979p96}. The reddening coefficient C(\hb) was derived in an iterative procedure assuming the same $\rm EW_{abs}$ for all the Balmer lines and the theoretical expected Balmer ratios at T$_{\rm e}$ = 7.6 $\times$ $10^3$~K \citep{Storey:1995p586}. Following this method, we find a coefficient of C(\hb) = 0.17 $\pm$ 0.03 and estimate the stellar absorption equivalent width to be always $\rm EW_{abs}\leq$2~\AA.

The dereddened emission line ratios with respect to $I(\hb)$ are given in Table~\ref{lambdaint}.  The corresponding errors are a combination of those derived in our profile-fitting procedure and the uncertainty in the continuum subtraction (\citealt{GonzalezDelgado:1994p555}). They do not include the uncertainty in the calibration that can be between 5 and 15 per cent, depending on the exact wavelength range (see Section 2.2). We find differences of $\leq$20 per cent with the emission-line ratios reported by  \citet{Esteban:2009p563}, except for the following lines: [\oii]$\lambda$3727, [\sii]$\lambda$6717, [\sii]$\lambda$6731 and H8+HeI for which we find differences of 40 per cent for [\oii]$\lambda$3727, 100 per cent differences for the [\sii] emission lines and 24 per cent for the last case. The differences for 
[\oii]$\lambda$3727 and H8+HeI can be attributed to the different spectral resolutions of both observations: 
\citet{Esteban:2009p563} are able to resolve the two [\oii] lines and the H8 and HeI separately, while the fluxes reported here are the combination of the unblended emission lines. The differences in the [\sii] emission lines are related to differences in the aperture used in each study: the slit field of view used by 
\citet{Esteban:2009p563} (5.76 $\times$ 1.7 arcsec$^2$) is small and it is located over a knot with high \ha\ surface brightness, while the fluxes reported here correspond to the total area of the region. Integrating over the slit field of view used in \citet{Esteban:2009p563}, we obtain differences of $\sim$10 per cent for both [\sii] emission lines. As we will see in Section~\ref{struction}, the [\sii] emission-line flux in each \emph{spaxel} shows a strong dependence on the same \emph{spaxel} \ha\ flux, which explains the differences found here with the [\sii] fluxes given in \citet{Esteban:2009p563}.

\begin{table}
\begin{center}
\caption{Dereddened emission-line ratios with respect to $I(\hb)$ detected in the integrated spectrum for the complete field of view of our observations (Fig.~\ref{espec_integ}). The errors are a combination of the profile-fitting procedure and the uncertainty in the continuum subtraction. f($\lambda$) is the reddening function normalised to \hb\ from \citet{Savage:1979p96}. }
\begin{tabular}{ccc}
\hline\noalign{\smallskip}
Emission Line &  $f(\lambda)$ &  $I(\lambda)$ \\  
\hline\noalign{\smallskip}

3727 $[$\oii$]$    &    0.252 &  2.86$\pm$ 0.05     \\     
3889 H8+HeI      &    0.220 &    0.12$\pm$ 0.02             \\     
3970 \hep\ + $[$\neiii$]$          &   0.204  &  0.11$\pm$ 0.02      \\ 
4102 \hdel\          &     0.177 &  0.21$\pm$ 0.01      \\  
4471 HeI              &   0.100 &  0.04$\pm$ 0.01 \\
4340 \hgam\         &    0.131 &  0.44$\pm$ 0.03     \\    
4959 $[$\oiii$]$     & -0.030 &  0.295$\pm$ 0.006         \\     
5007 $[$\oiii$]$     & -0.043 &  0.95$\pm$ 0.01          \\    
5876 He~I            &   -0.220  &  0.099$\pm$ 0.003       \\     
6563 \ha\             &  -0.314   &  2.94$\pm$ 0.02          \\
6584 $[$\nii$]$     &  -0.319 &  0.505$\pm$ 0.005         \\   
6678 HeI             &   -0.329 &  0.032$\pm$ 0.003   \\ 
6717 $[$\sii$]$    &  -0.335  &  0.267$\pm$ 0.005          \\           
6731 $[$\sii$]$     &  -0.336 &  0.186$\pm$ 0.004          \\        
\hline\noalign{\smallskip}
\end{tabular}
\label{lambdaint}
\end{center}
\end{table}

In Table~\ref{tabint}, we show the principal diagnostic emission-line ratios derived from the integrated spectrum as well as the physical parameters derived from them. The integrated reddening coefficient corresponds to an extinction of 0.36 mag at \ha, which agrees with the values reported in \citet{Relano:2009p558} derived from other methods. 

Due to the contribution of a strong sky HgI4358 emission line and the low intensity of [\oiii]$\lambda$4363 we were not able to detect this emission line, and thus a measurement of the temperature with these observations cannot be made. We report here the mean value obtained from \citet{Esteban:2009p563} using high-resolution echelle spectroscopy, which has been used to obtain the electron density from the [\sii]$\lambda$6717/[\sii]$\lambda$6731 emission line ratio. We derived a nominal density value of
13~\cmtres, for a line ratio of 1.43 (see Table~\ref{tabint}), which agrees with the low density scenario derived by \citet{Esteban:2009p563}. Taking into account the extinction-corrected \ha\ luminosity of NGC~595 reported in \citet{Relano:2009p558} and the electron densities derived here, we estimate an ionization parameter of q$_{\rm eff}$=1.9$\times 10^7$~cm~s$\me$ for an electron density face value of n$_{\rm e}$=20 \cmtres. 

The metallicity of the \hii\ region can be derived using information from several emission-line ratios. For this purpose, we will use as metallicity calibrator the widely used R$_{\rm 23}$ index \citep{Pagel:1979p549}. 
As we will discuss later in Section~\ref{metstruc}, this parameter is substantially constant across the face of the region, and thus we use it here to derive a representative value for the \hii\ region metallicity. The calibration of R$_{\rm 23}$ with metallicity has been extensively studied in the literature, either using photoionization models (e.g. \citealt{McGaugh:1991p514}; \citealt{Kewley:2002p405}) or from empirical calibrations (e.g. \citealt{Pilyugin:2000p550}; \citealt{Pilyugin:2001p94}). One of the major problems of R$_{\rm 23}$ is that it is double valued; thus an independent metallicity diagnostic must be used in order to set the region in the upper or lower branch in the Z-R$_{\rm 23}$ diagram.  
We use the observed [\nii]/\ha\ line ratio to make a first estimation of Z: following the relation given by \citet*{Denicolo:2002p548} we obtain 
12 + log[O/H] = 8.56, while the calibration of \citet{Pettini:2004p551} gives 12+log[O/H]=8.41. Both results shows that we are able to use the upper branch of the Z-R$_{\rm 23}$ diagram. We apply two calibrations to obtain Z from R$_{\rm 23}$, 
 \citet{McGaugh:1991p514} as parametrised by \citet*{Kobulnicky:1999p541}, which relies on photoionization models, and the empirical calibration of \citet{Pilyugin:2001p94}. The first one gives 12 + log[O/H] = 8.75, while the empirical relation yields 12+log[O/H]=8.43. The mean value between these two calibrations, 12+log[O/H]=8.59, was finally chosen to characterize the oxygen abundance for the whole \hii\ region, with a typical uncertainty of $\pm$0.32~dex taking into account the different results from both calibrations. This value agrees with the range of previous values given in the literature: 8.45 $\pm$ 0.03 (t$^2$ = 0.000) and 8.69 $\pm$ 0.05 (t$^2$ = 0.036) by \citet{Esteban:2009p563} and 8.44$\pm$0.09 by \citet{Vilchez:1988p505}.

\begin{table}
\begin{center}
\caption{Principal diagnostic emission-line ratios and integrated physical properties of NGC~595. Except for the electronic temperature that was taken as a mean value of the temperatures derived by \citet{Esteban:2009p563}, the rest of the physical properties have been 
estimated here using our integrated fluxes (see text for details).\rdostres = ([\oii] $\lambda$3727 +  [\oiii] $\lambda\lambda$4959,5007)/\hb, as defined by \citet{Pagel:1979p549}.}
\begin{tabular}{cc}
\hline\noalign{\smallskip}
Parameter &  Value \\  
\hline\noalign{\smallskip}
log $[$\nii$]$6584/\ha\                            &   -0.765$\pm$0.005   \\
log $[$\sii$]$(6717+6731)/\ha\                &   -0.813$\pm$0.007 \\
log $[$\oiii$]$$\lambda$(4959+5007)/$[$\oii$]$$\lambda$3727 & -0.361$\pm$0.009   \\
log R$_{\rm 23}$                      &  0.614$\pm$0.006   \\
$[$\sii$]$$\lambda$6717/6731         &  1.43$\pm$0.04   \\
\hline
  C(\hb)                                                      &  0.17$\pm$0.03 \\
  n$_{\rm e}$ (\cmtres)                               &  $<$220 \\
  q$_{\rm eff}$ (cm~s$\me$)                  &  1.9$\times 10^7$  \\
  T$_{\rm e}$ (K)                              & 7670$\pm$116 (adopted) \\
  12+log[O/H]                              &  8.59$\pm$0.32 \\
  \hline\noalign{\smallskip}
\end{tabular}
\label{tabint}
\end{center}
\end{table}

\section{Emission line maps}
\subsection{Structure of the ionized gas and the stellar component}

\begin{figure}
\begin{center}
\begin{minipage}[h]{0.35\textwidth}
\includegraphics[width=\textwidth]{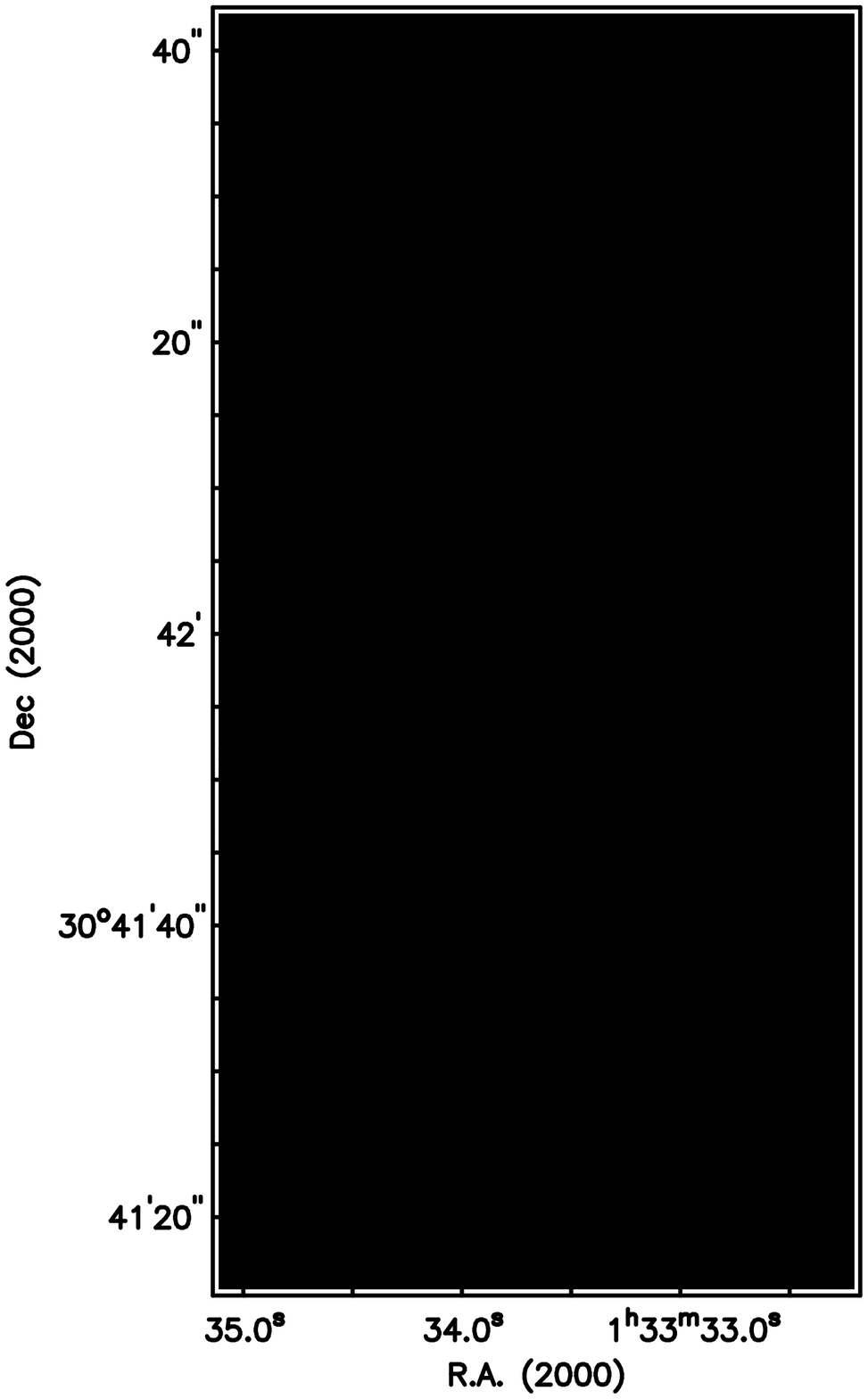}
\end{minipage}
\caption{Map of the observed \ha\ flux derived using the fitting procedure described in Section~2.3. The field of view of the observations is depicted in Fig.~\ref{apuntado}. Each \emph{spaxel} has a 1 $\times$ 1 arcsec$^2$ size and fluxes are in arbitrary units. The contours correspond to the WFPC2/F336W filter image obtained from the HST MAST Archive. The most intensity contours show the location of the ionizing stellar clusters within the region.}
\label{flux_ha_hb}
\end{center}
\end{figure}

In Fig.~\ref{flux_ha_hb}, we show a map of the observed \ha\ flux for NGC~595 derived with the procedure explained in 
Section~\ref{line_fit}. The similarity with the observed  \ha\ flux obtained from direct image (see Fig.~\ref{apuntado}) shows the power of the IFS observations to reproduce the surface brightness distribution for this particular region. The \ha\ shell structure is clearly delineated, it extends from the south-west to the north in an arched structure with \ha\ maxima located close to the ionizing stars. The separation between the central star clusters and the  \ha\ maxima is 
$\sim$6 arcsec, $\sim$22~pc. The external low \ha\ surface brightness shell, seen in Fig.~\ref{apuntado} eastwards to the brightest one, is also observed in the  \ha\ Êmap, which shows the high quality of the data to trace the diffuse  \ha\ emission of the region. In Fig.~\ref{flux_ha_hb}, we overplot emission contours of the Wide Field Planetary Camera 2 (WFPC2)/F336W image from the {\it Hubble Space Telescope (HST)} Multimission Archive at the Space Telescope Science Institute (MAST) Archive. The knots correspond to the location of the stellar clusters; two of the most intense ones are located in the centre of the shell and do not correspond to regions of high \ha\ emission; the third intense knot is within the \ha\ shell structure, slightly north-east from the others. Farther away from the most intense knots, there are others spatially correlated with diffuse \ha\ emission. 

\subsection{Extinction in  NGC~595}

\label{extinc_section}
\begin{figure*}
\begin{minipage}[]{0.75\textwidth}
\includegraphics[width=0.5\textwidth]{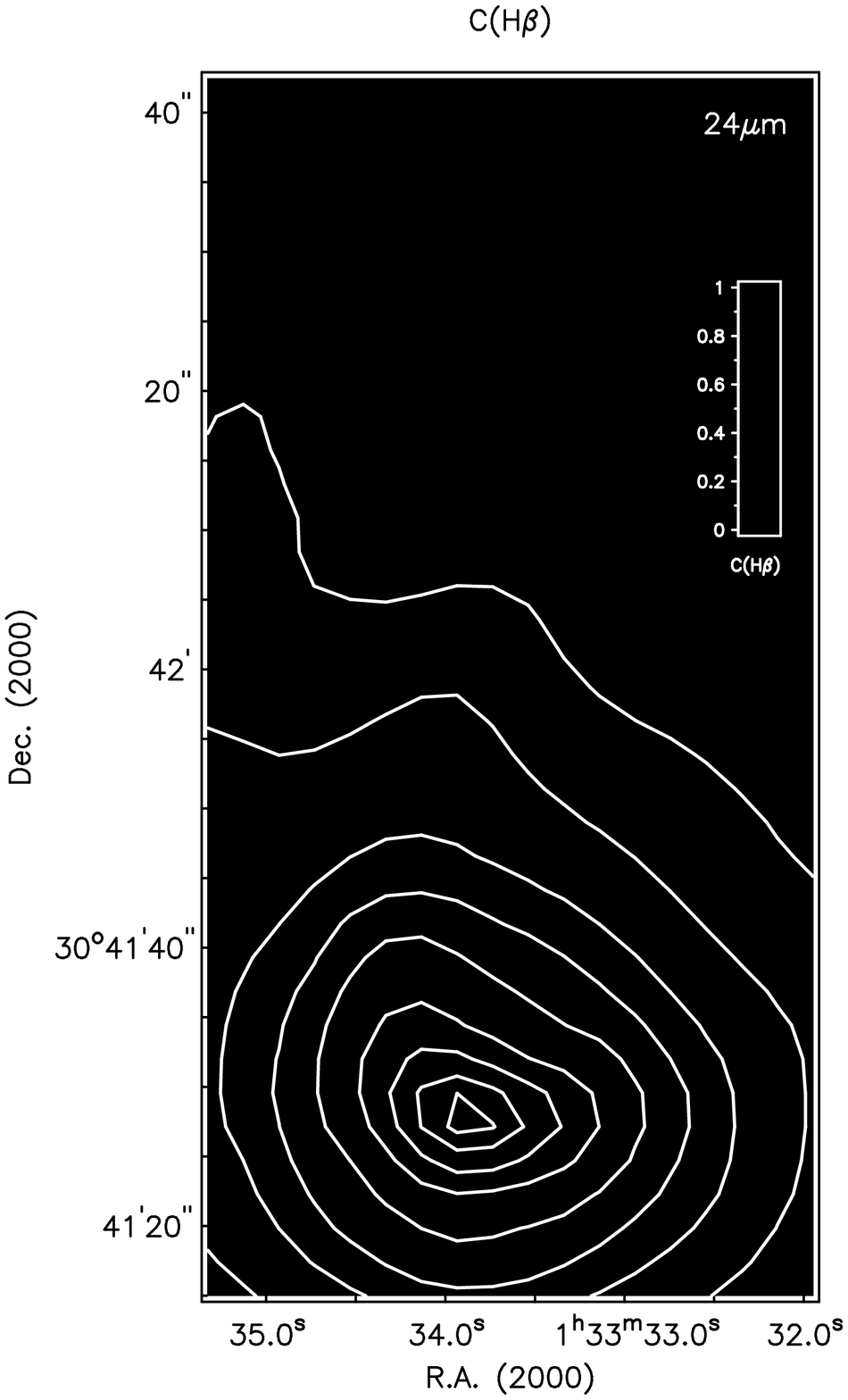}
\includegraphics[width=0.5\textwidth]{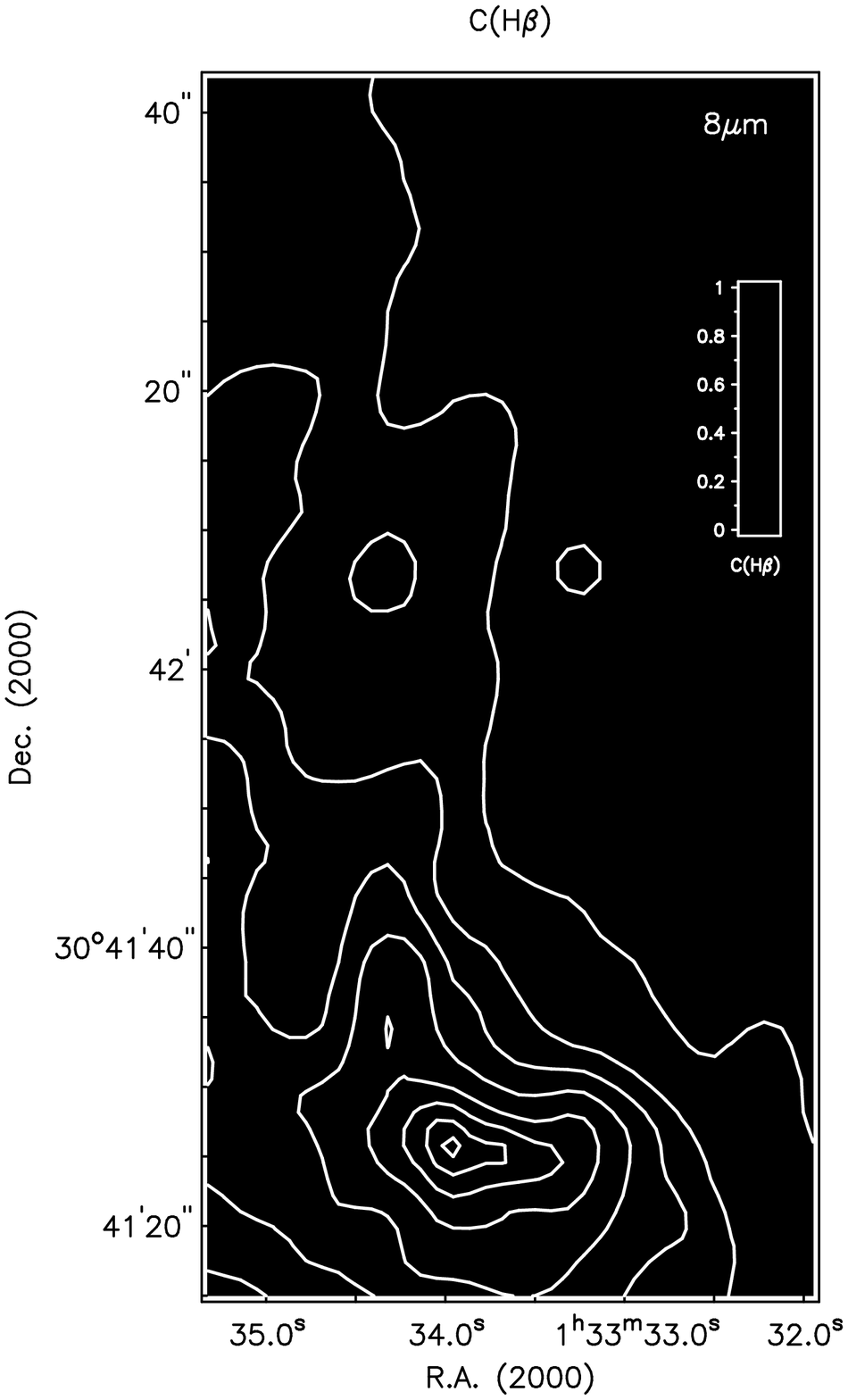}  
\end{minipage}

\begin{minipage}[]{0.75\textwidth}
\includegraphics[width=0.5\textwidth]{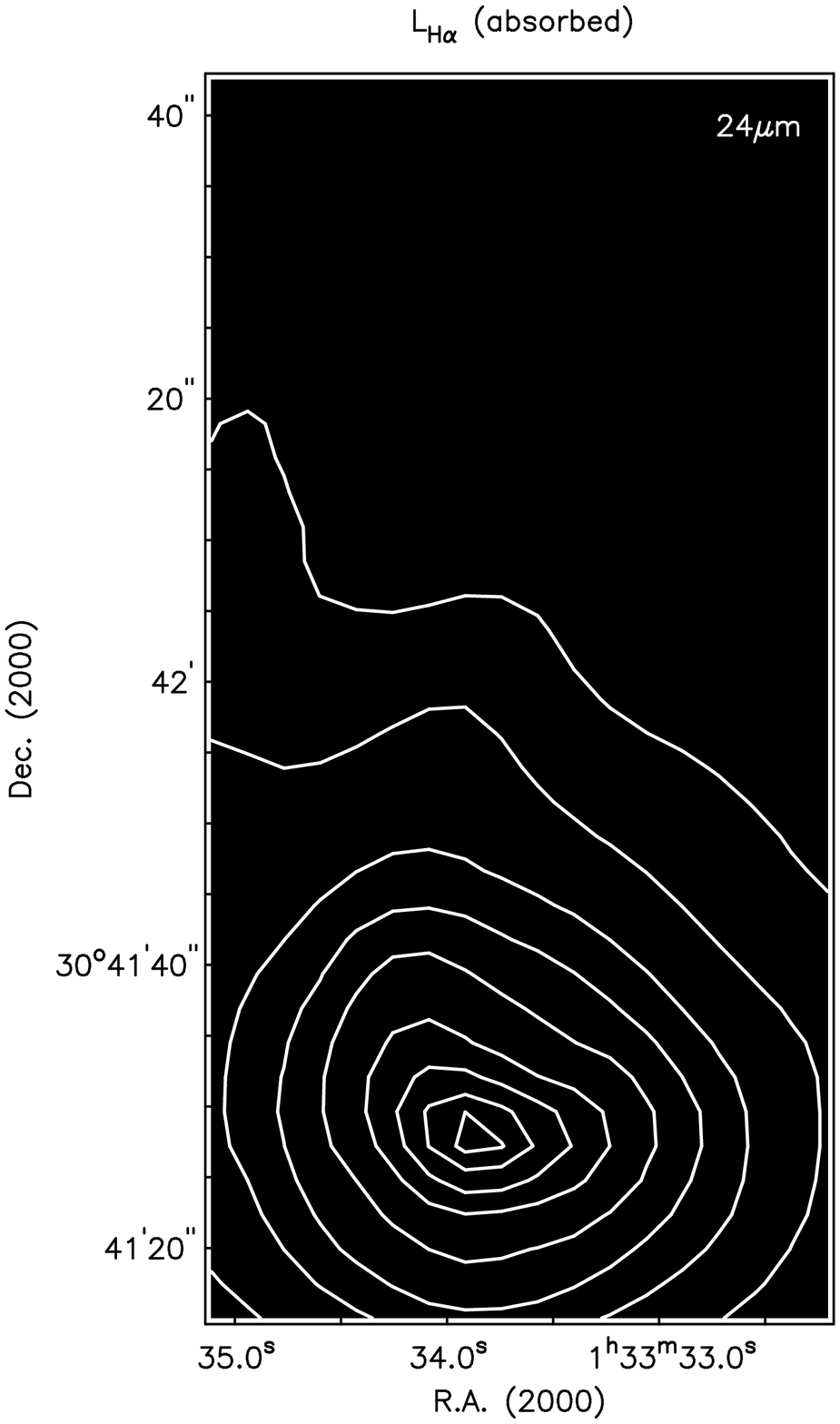}
\includegraphics[width=0.5\textwidth]{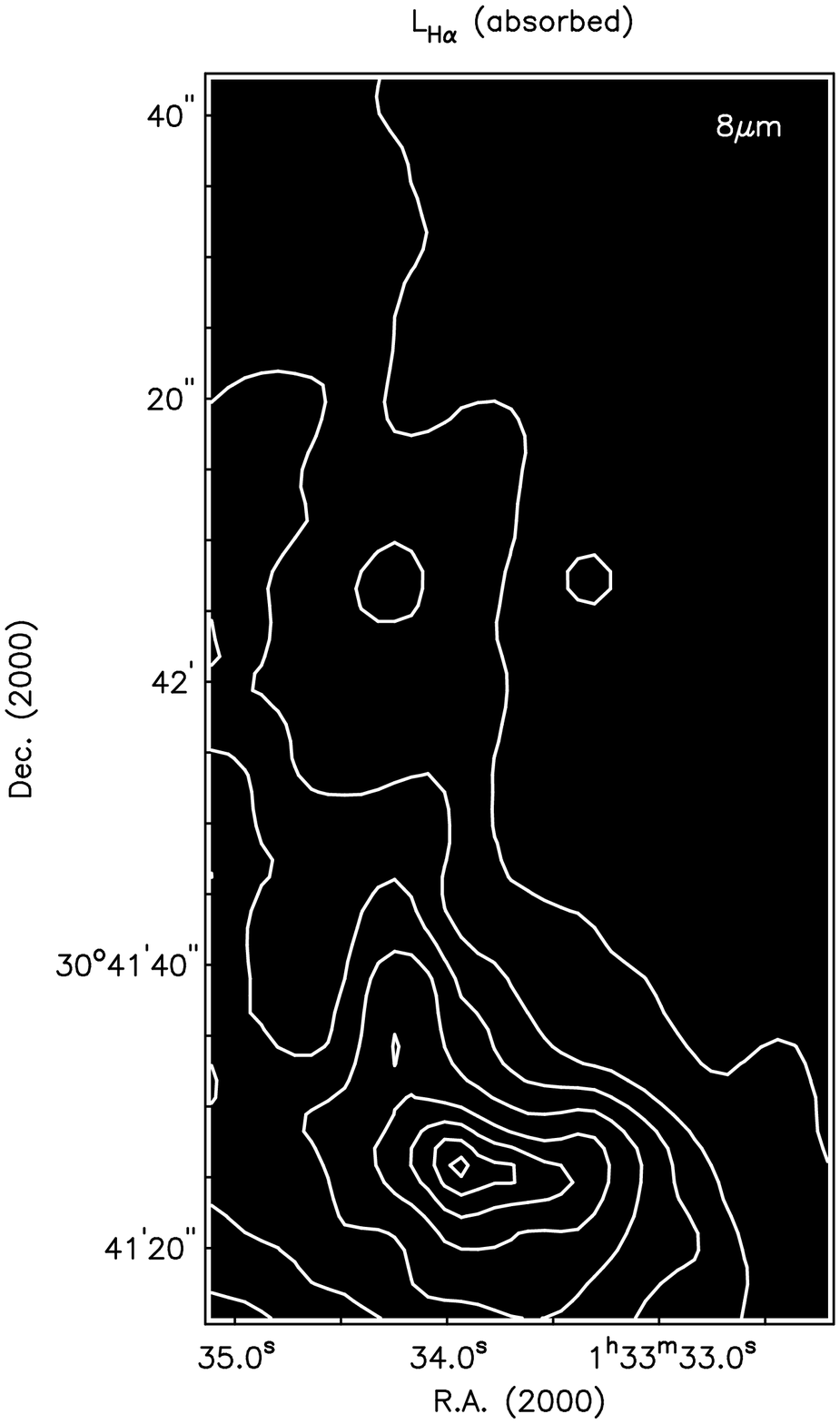}  
\end{minipage}

\caption{Upper: Map of the reddening coefficient C(\hb) for NGC~595 with 24~\mi\ (left) and 8~\mi\ (right) emission contours overplotted. 
The intensity contours for the 24 and 8~\mi\ emission are at (2, 5, 10, 20, 40, 60, 80, 95) per cent of the maximum intensity within the region.  
A 1 per cent contour level corresponds to 3$\sigma$ and 1$\sigma$ for the 24 and 8~\mi\ emissions, respectively. Lower: {\it Absorbed} \ha\ luminosity of NGC~595 with 24~\mi\ (left) and 8~\mi\ (right) emission contours overplotted. The contour levels are the same as in the upper figures.}
\label{ext+ir}
\end{figure*}

We used the observed \ha/\hb\ ratio to derive a map of the reddening coefficient C(\hb) and to correct the emission-line fluxes at each \emph{spaxel} of the field of view. The ratio was compared with the theoretical expected value of the Balmer decrement for case B of recombination theory at T$_{\rm e}$ = 7.6 $\times$ $10^3$~K \citep{Storey:1995p586}, and then by using the extinction law of \citet{Savage:1979p96} we obtained C(\hb). The Balmer emission lines can be affected by the absorption of the underlying stellar component, which would affect the derived values of the reddening coefficient. We explore the influence of this effect in our data by visual inspection of the spectra and by a
comparison of the observed \hb\ equivalent width (EW(\hb)) in emission and the expected absorption equivalent width (see Section~\ref{secint}). 
Except for a circular region of $\sim$3~pixels associated with the location of the central clusters where we measure EW(\hb) of $\sim$5-10~\AA, we find EW(\hb) within a range of 40--450~\AA. The effect of the underlying stellar population is negligible at the locations of the high equivalent widths. A closer inspection of the Balmer emission-line profiles at the location of the central clusters would suggest the presence of absorption signatures for \hdel\ and \hep\ which are marginally detected (S/N$\leq$3) and consistent with $\rm EW_{abs}\sim$(1.0-1.4)~\AA.

A C(\hb) map is shown in the upper panels in Fig.~\ref{ext+ir}. It presents a very concentrated distribution within the region, similar to the extinction structure obtained by \citet{Viallefond:1983p504} from radio and \ha\ observations. We derive values for the extinction at \ha\ in the centre of the region within the range of 1.0-1.5~mag, which agrees with the range reported in \citet{Viallefond:1983p504} for the Balmer extinction in the core of the nebula, 
(0.88-1.15)~mag. With the reddening coefficient map we can then correct the observed spectrum in each \emph{spaxel} within the region and obtain maps of the emission-line ratios that can be used to study the physical properties of the \hii\ region. As an illustration, we show in Fig.~\ref{lineratio} some of the emission-line ratio maps that will be analysed later in this paper. 

In Fig.~\ref{ext+ir} (upper panels) we overplot contours of the 24 and 8~\mi\ emissions from {\it Spitzer} (\citealt{Relano:2009p558}). The location where C(\hb) is high corresponds within the spatial resolution ($\sim$1.5 arcsec for our observations and $\sim$6 arcsec for the observations at 24~\mi)
to high values of the 24~\mi\ emission. The emission at 8~\mi\ also correlates to the maximum values of C(\hb), but it is not as concentrated as the emission at 24~\mi\ (Fig.~\ref{ext+ir}, upper left). In the lower panels of Fig.~\ref{ext+ir}, we show the relation of the emission at 24 and 8~\mi\ with the {\it absorbed} \ha\ luminosity of the region. 
The \ha\ luminosity absorbed in the region is obtained as the difference between the total extinction-corrected \ha\ luminosity and the \ha\ luminosity corrected for the foreground Galactic extinction (E(B-V)=0.041 \citep*{Schlegel:1998p565} is used to account for the Galactic extinction). As it is shown in Fig.~\ref{ext+ir} (lower left panel), the 24~\mi\ emission spatially correlates with the {\it absorbed} \ha\ luminosity, with the maximum at 24~\mi\ coinciding with the central position of the {\it absorbed} \ha\ structure. The spatial 
correlation of the 8~\mi\ emission and the {\it absorbed} \ha\ luminosity is not so strong; the maximum at 8~\mi\ is shifted outwards of the {\it absorbed} \ha\ shell structure. 

\begin{figure}
\begin{center}
\begin{minipage}[h]{\textwidth}
\includegraphics[width=0.5\textwidth]{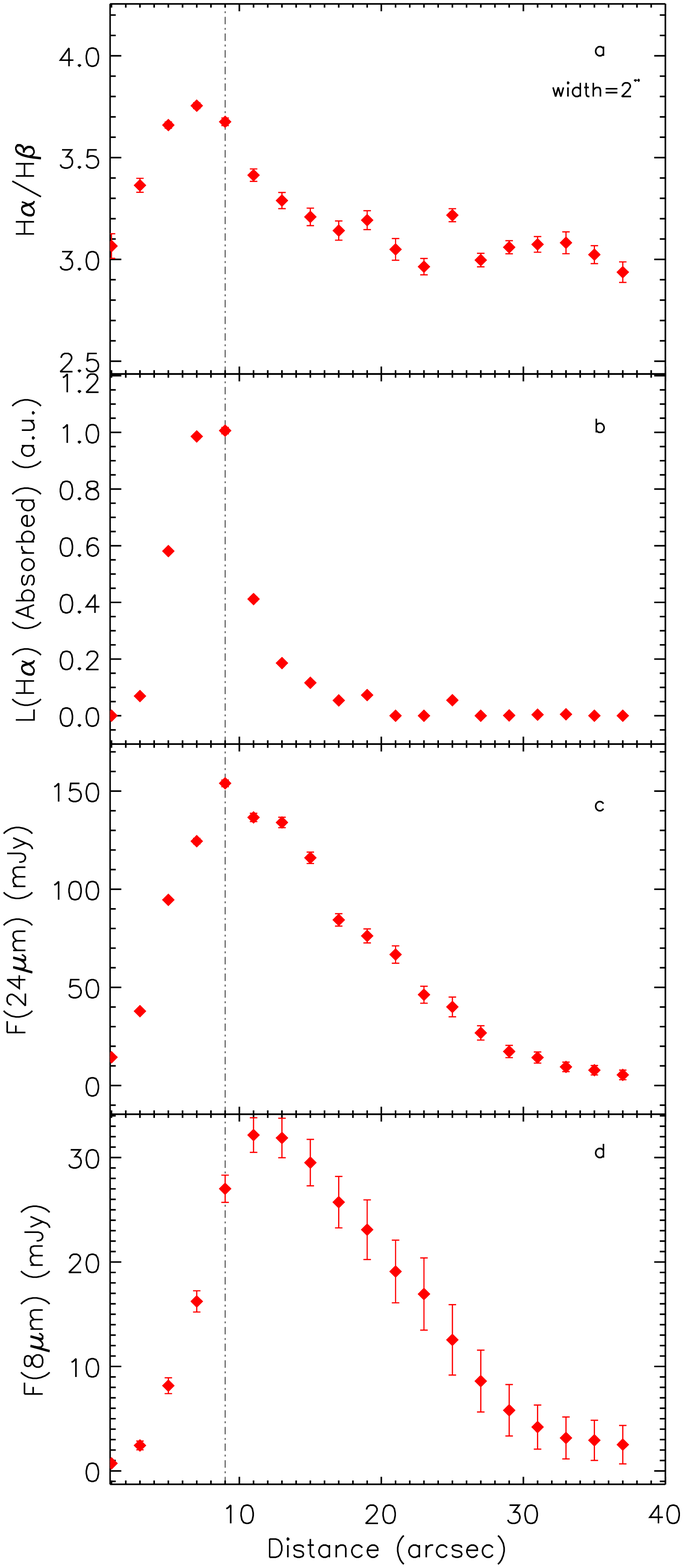}
\end{minipage}
\vspace{-2.5cm}
\caption{Radial profiles (from top to bottom) of the (a) \ha/\hb\ emission line ratio, (b) {\it absorbed} \ha\ luminosity  and {\it Spitzer} infrared bands at (c) 24~\mi\ and (d) 8~\mi. For panels (a) and (b), we isolated the signal coming from concentric elliptical rings of 2 arcsec width projected onto the mayor axis of the ellipse and derived the corresponding integrated spectrum at each annulus. The center of the ellipse (-2,0) is expressed in the units of the emission-line maps shown in Fig.~\ref{lineratio}. We fitted the spectra in the same way as explained in Section~\ref{line_fit} and derived the integrated emission-line fluxes for each concentric annulus (see text). For the  24 and 8~\mi\ emission profiles, we plot the integrated fluxes for each elliptical annulus. The dot-dashed line represents the location of the maximum of the {\it absorbed} \ha\ luminosity.}
\label{perfextratio}
\end{center}
\end{figure}

\begin{figure*}
\begin{minipage}[]{0.75\textwidth}
\includegraphics[width=0.5\textwidth]{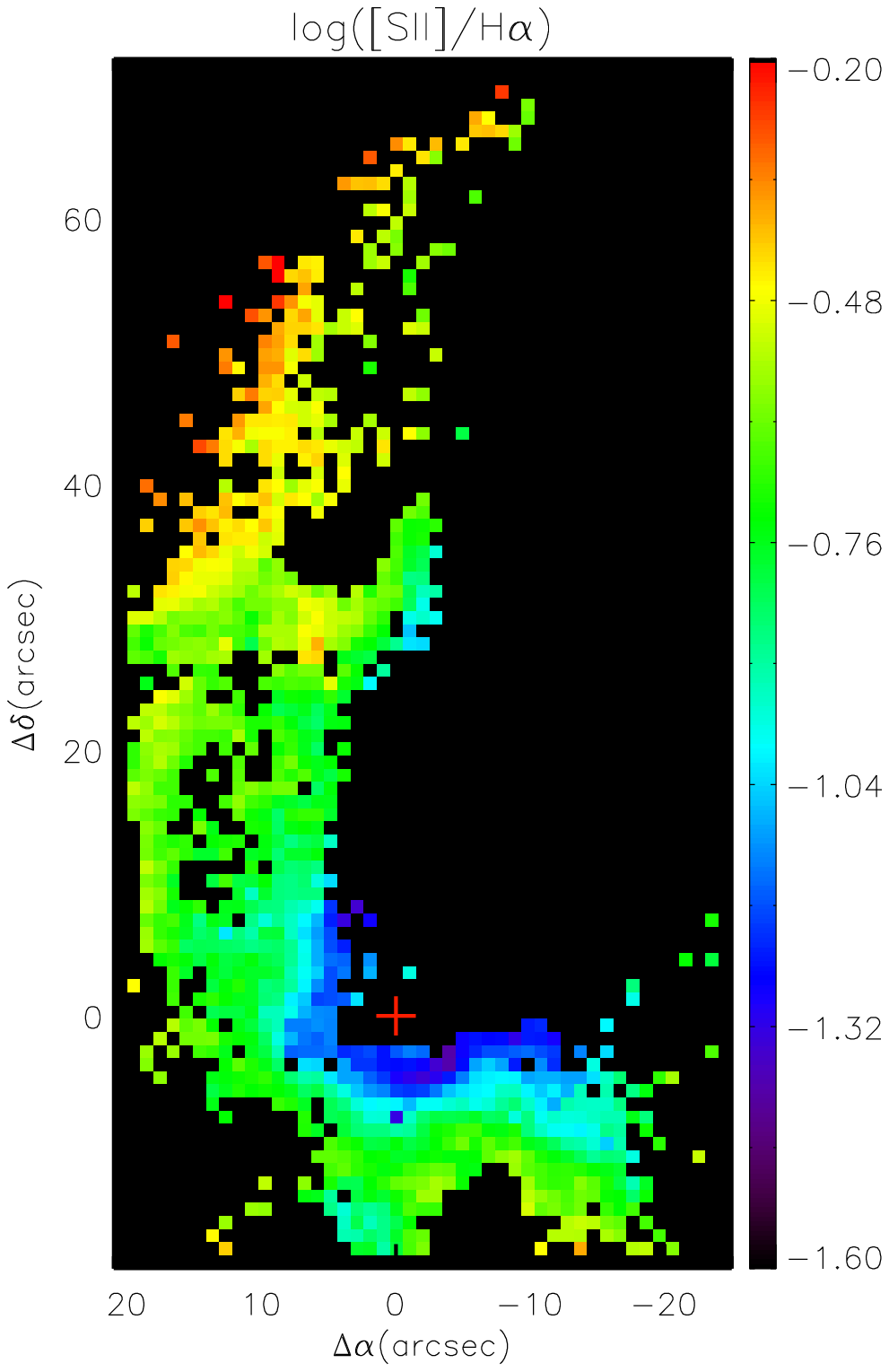}
\includegraphics[width=0.5\textwidth]{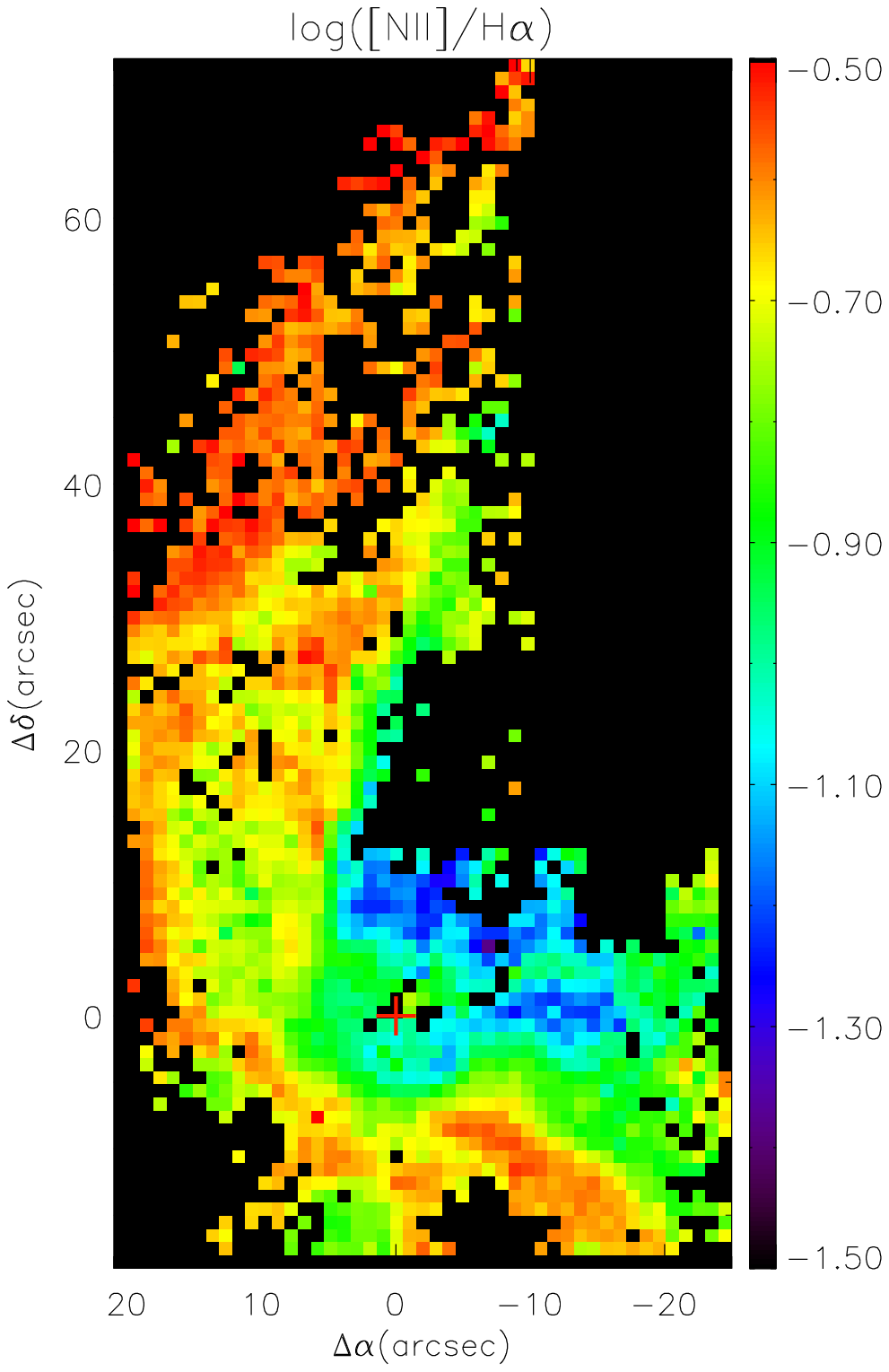} 
\end{minipage}
\vspace{-2.0cm}

\begin{minipage}[]{0.75\textwidth}
\includegraphics[width=0.5\textwidth]{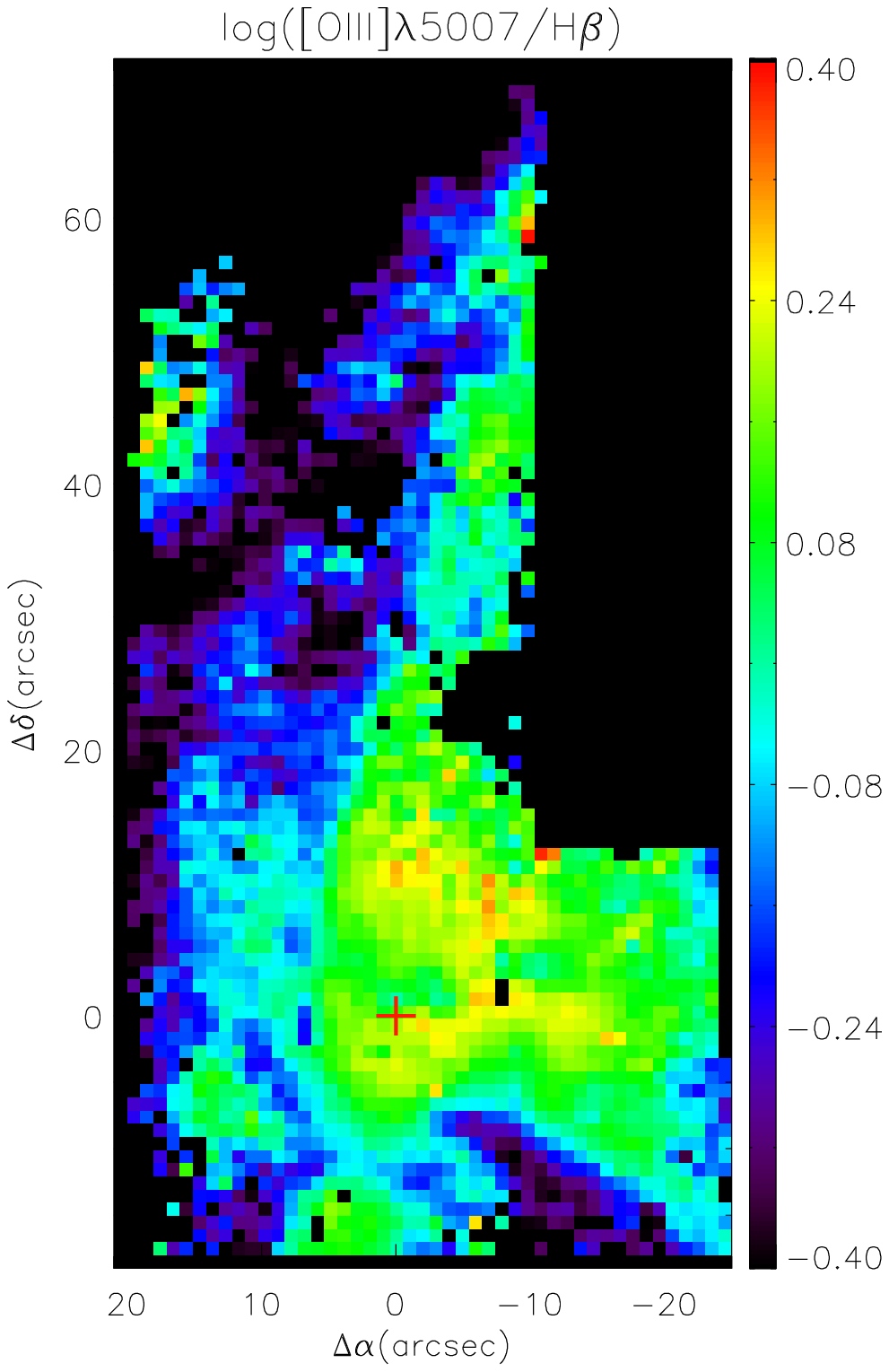}
\includegraphics[width=0.5\textwidth]{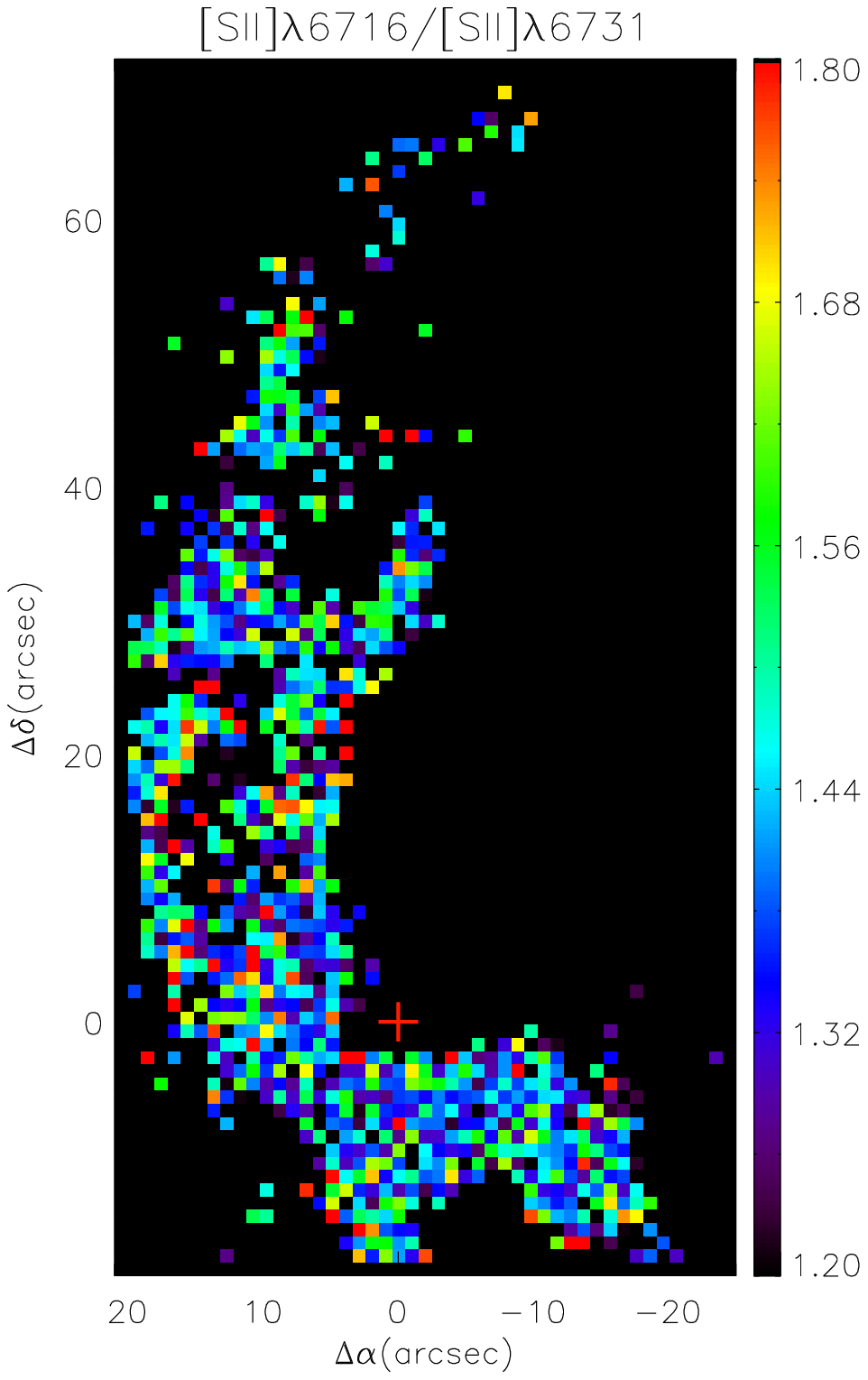}
\end{minipage}
\vspace{-1.0cm}
\caption{[\sii]$\lambda$6717,31/\ha\ (upper left), [\nii]$\lambda$6584/\ha\ (upper right), [\oiii]$\lambda$5007/\hb\ (bottom left) and [\sii]$\lambda$6717/[\sii]$\lambda$6731 (bottom right) emission-line ratio maps for the whole face of NGC~595. Only \emph{spaxels} with emission-line ratios having relative errors $<$ 30 per cent are shown. Extinction correction for the fitted emission lines was performed in each \emph{spaxel} prior obtaining the emission-line ratios shown here. The red cross marks the location of the central and most intense stellar cluster
(R.A. (J2000): 1h~33m~33.79s, DEC(J2000): 30d~41m~32.6s) and distances are relative to this position. The orientation is north up, east to the left.}
\label{lineratio}
\end{figure*}

These spatial correlations are better shown in Fig.~\ref{perfextratio}, where we plot the emissions integrated in elliptical concentric annuli as a function of the radial distance measured along the major axis of an ellipse centred at the location of the main stellar clusters (we take this position to be 2 arcsec westwards from the most intense stellar cluster marked as a red cross in the emission-line maps of Fig.~\ref{lineratio}). The major to minor axis ratio of the ellipse is derived using the inclination angle of the galaxy (i=56\degree\ for M33 \citet{vandenBergh:2000p502}), and we choose a position angle of 129\degree\ for the major axis since this orientation better traces the shell structure of the region. In this configuration, we use rings of 2 arcsec widths to obtained the elliptical profiles. 

The \ha/\hb\ emission line ratio was derived for each annulus in the original data cube. The resulting spectrum for each annulus was fitted in the same way as we explained in Section~\ref{line_fit}, and the corresponding fluxes at \ha, \hb\ were used to obtain the radial profiles shown in Fig.~\ref{perfextratio}(a). The {\it absorbed} \ha\ luminosity profile (Fig.~\ref{perfextratio}b) was obtained in the same way we derived the {\it absorbed} \ha\ luminosity map. 
 
The comparison of Fig.~\ref{perfextratio}(a) and (c) shows that the radial distribution of the \ha/\hb\ ratio and the 24~\mi\ emission are similar, with the same radial decline towards larger radial distances and maxima slightly shifted ($\sim$2 arcsec, lower than the resolution of the 24~\mi\ observations ($\sim$6 arcsec)). The comparison of Figs.~\ref{perfextratio}(b) and (c) shows that the distribution of the {\it absorbed} \ha\ luminosity and the 24~\mi\ emission are also similar, with their respective maxima located at the same radial distances. The spatial relations shown here between the 24~\mi\ emission, the extinction suffered by the ionized gas and the  {\it absorbed} \ha\ luminosity support the idea that the 24~\mi\ emission is produced by heating of dust probably mixed with the ionized gas inside the region, which was previously suggested by \citet{Relano:2009p558}. 

The emission at 8~\mi\ also correlates to the maximum values of C(\hb), but the spatial distribution of the 8~\mi\ emission does not follow the C(\hb) structure as the 24~\mi\ emission distribution is (Fig.~\ref{ext+ir}, upper panels). In Fig.~\ref{perfextratio}(d), we plot the 8~\mi\ emission elliptical profiles. The maximum at 8~\mi\ is $\sim$4-5 arcsec shifted towards higher radial distances than the maximum of the \ha/\hb\ ratio. This shows that the dust responsible for the Balmer extinction is the dust emitting at 24~\mi\ rather than the dust emitting at 8~\mi. \citet{Relano:2009p558} showed that the 8~\mi\ emission for NGC~595 is more related to the location of the main CO molecular clouds identified in this region than to the position of the ionized gas. 

\subsection{Density Structure}
\label{density}

\begin{figure}
\begin{center}
\begin{minipage}[h]{\textwidth}
\includegraphics[width=0.5\textwidth]{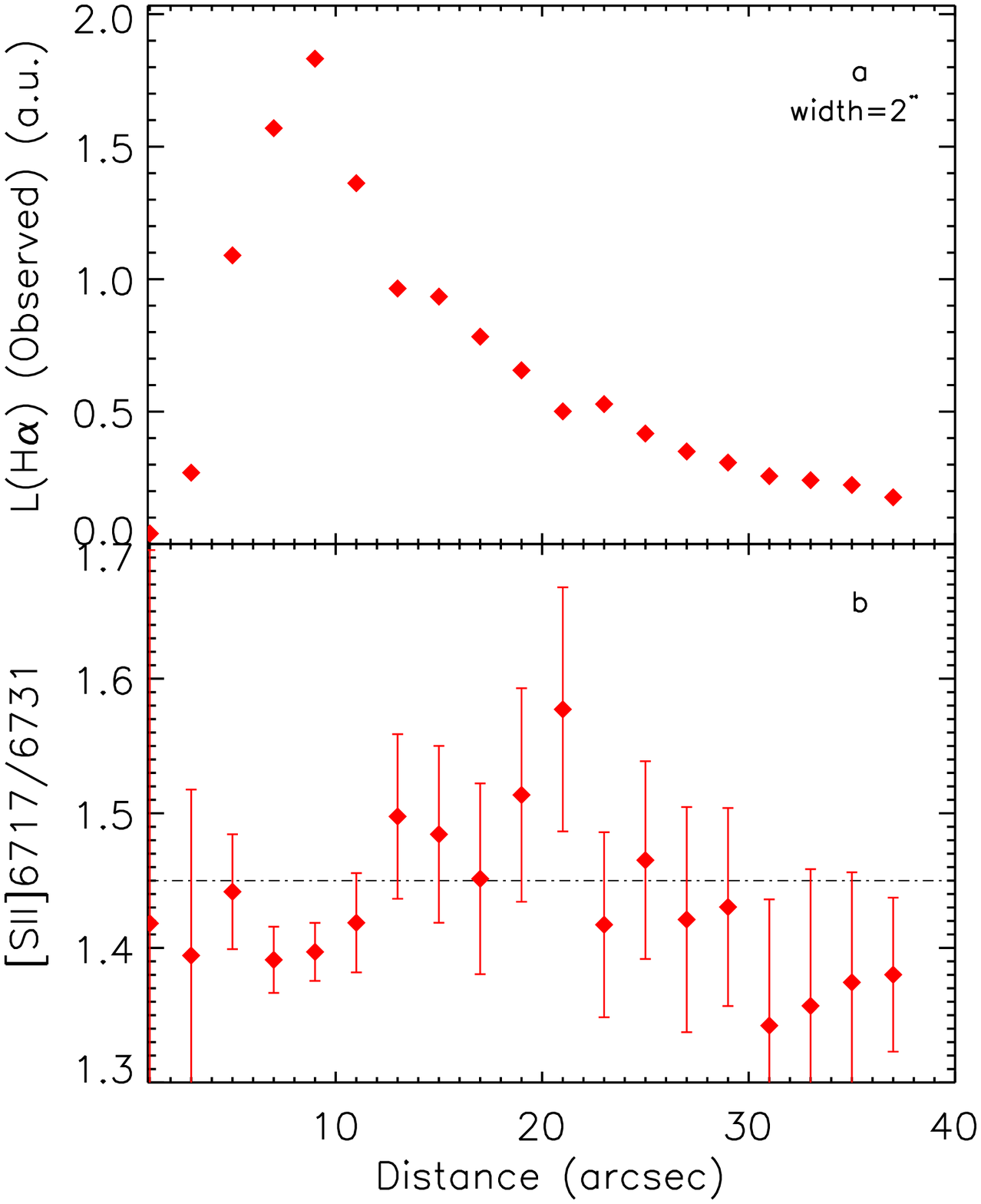}
\end{minipage}
\caption{Radial profiles of the observed \ha\ luminosity (a),  [\sii]$\lambda$6717/[\sii]$\lambda$6731 (b), obtained in the same way as the profiles shown in Fig.~\ref{perfextratio}.}
\label{perfsiiratio}
\end{center}
\end{figure}

We study the electron density using the [\sii]$\lambda$6717/[\sii]$\lambda$6731 emission-line ratio. A map of this ratio is shown in Fig.~\ref{lineratio} (bottom-right). The emission-line ratio ranges from 1.2 to 1.8 (n$_{\rm e}<$~220~\cmtres), showing that the \hii\ region has in general a low electron density, as was also found by \citet{Esteban:2009p563} and \citet{Vilchez:1988p505}. The map shows no particular density structure, and further maps obtained from binning the data cube in 2$\times$2 and 4$\times$4 \emph{spaxels} do not show any density structure neither. \citet{Lagrois:2009p564} show a density map for NGC~595 which extends in a wider field of view than ours; the lower density values in their map are associated to 
the location of HI neutral gas, while within the area cover by \ha\ emission their map shows no strong variations with density values of $\sim$60-150\cmtres. 

However, the pronounced \ha\ shell morphology of the nebula would suggest a density enhancement at the location of the \ha\ maxima. Density variations within the \hii\ regions have been reported before and in some cases, a relation between the density and the \ha\ emission has been observed \citep*{Castaneda:1992p479}. In order to check carefully any possible density variation with the \ha\ emission, we have compared the radial profiles of the [\sii]$\lambda$6717/[\sii]$\lambda$6731 line ratio (Fig.~\ref{perfsiiratio}~b) with profiles of observed \ha\ emission (Fig.~\ref{perfsiiratio}~a). 

The mean value for the [\sii]$\lambda$6717/[\sii]$\lambda$6731 line ratio in all the rings is (1.43$\pm$0.06), corresponding to the low density limit. The elliptical profiles show that there is no electron density variation with the \ha\ luminosity distribution for this region. In this situation, at the location of the maximum in \ha\ luminosity, which would correspond to an enhancement of mean electron density, there should be an increase of the filling factor producing a higher emission measure along the line of sight. In order to further study this effect, detailed  photoionization models of the region would need to be performed (P\'erez-Montero et al., in preparation).

\subsection{ionization structure}
\label{struction}

Emission line ratio maps of [\sii]$\lambda$6717,31/\ha, [\nii]$\lambda$6584/\ha, [\oiii]$\lambda$5007/\hb\ and 
[\sii]$\lambda$6717/[\sii]$\lambda$6731 are shown in Fig.~\ref{lineratio}. The last ratio gives information of the density of the \hii\ region and was studied in previous section; the others allow us to study the ionization structure of the region. Lower values of [\sii]$\lambda$6717,31/\ha\ and [\nii]$\lambda$6584/\ha\ are located close to the position of the ionizing stars, corresponding to the high excitation zone within the region, while at the outskirts the values of these ratios are higher, depicting the low excitation zone. The same effect is seen in the [\oiii]$\lambda$5007/\hb\ map; the high excitation zone closer to the central stars is shown by the high values of this ratio, while the low excitation zone is observed far from the central region, corresponding to low values of [\oiii]$\lambda$5007/\hb.

\begin{figure*}
\includegraphics[width=\textwidth]{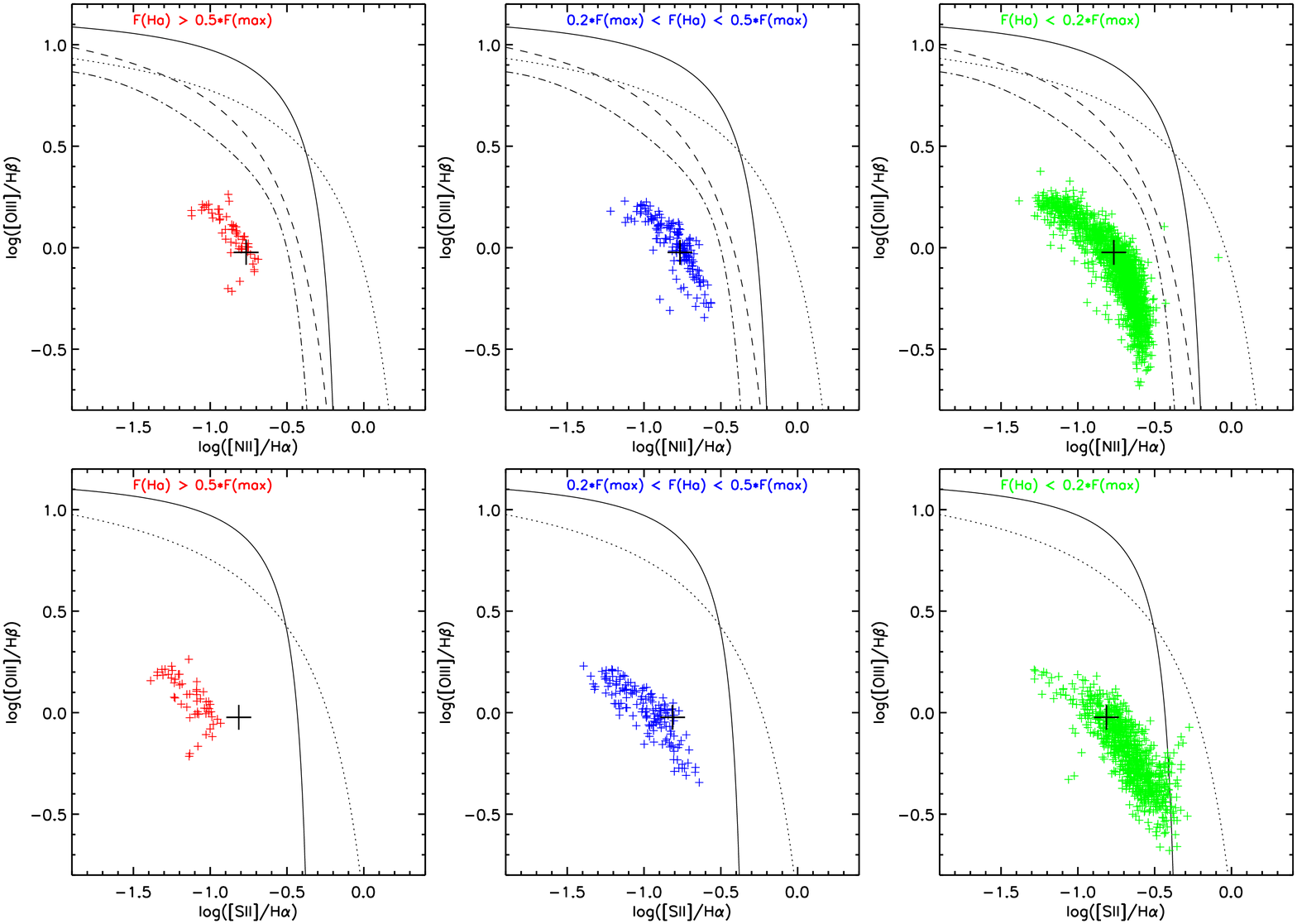}
\caption{BPT emission-line ratio diagnostics \citep{Baldwin:1981p100,Veilleux:1987p477} for all the \emph{spaxels} in the field of view where we measure [\oiii]/\hb, [\nii]/\ha\ and [\sii]/\ha\ in bins of different \ha\ flux. The black cross is the value obtained using the total integrated fluxes of the \hii\ region. The lines show the separation between AGN and NSF galaxies from different studies: continuous line -- \citep{Veilleux:1987p477}, dot line -- \citep{Kewley:2001p119}, dashed line -- \citep{Kauffmann:2003p513}, and dashed-dot line -- \citep{Stasinska:2006p475}.}
\label{BPTdiag}
\end{figure*}

Using BPT diagrams \citep*{Baldwin:1981p100,Veilleux:1987p477} and the emission-line ratio maps in Fig.~\ref{lineratio} we can study whether non-photoionizing mechanisms (e.g. shocks) are important within NGC~595. In Fig.~\ref{BPTdiag} we show the BPT diagrams for all the \emph{spaxels} in the field of view with measurements of [\oiii]/\hb, [\nii]/\ha\ and [\sii]/\ha\ grouped into bins of \ha\ flux. We also show the separation between active galactic nuclei (AGN) and normal star-forming (NSF) galaxies from different studies \citep{Veilleux:1987p477,Kewley:2001p119,Kauffmann:2003p513,Stasinska:2006p475}. As can be seen in the figure, all the \emph{spaxels} lie below the separation lines, which shows that the main mechanism producing the observed emission lines within NGC~595 is probably photoionization. We have compared (not shown here) the location of the \emph{spaxels} in these diagrams with the one for a wide sample of \hii\ regions from the literature. The area occupied by the set of \hii\ regions from the literature could be considered as the representative location for \hii\ regions, where photoionization dominates. The comparison shows that all our \emph{spaxels} fall into this area; therefore, if non-photoionizing mechanisms are present in the region, these are not contributing substantially to the emission of NGC~595. However, a detailed comparison of these observed line ratios for some \emph{spaxels} with the predictions of models by \citet{Allen:2008p562} suggests that somewhat localized effects of shocks can not be disregarded. 

In the top panels of Fig.~\ref{BPTdiag} we study the [\oiii]/\hb\ versus [\nii]/\ha, while the bottom panels show [\oiii]/\hb\ versus [\sii]/\ha. The distribution of the data is different in each set of panels.  
In the case of [\oiii]/\hb\ versus [\nii]/\ha, the \emph{spaxels} with a high \ha\ flux (top-left hand panel) correspond to those showing high excitations levels (high values of [\oiii]/\hb); however, the low \ha\ flux \emph{spaxels} (top-right hand panel) cover the whole range of excitations within the nebula. The highest [\oiii]/\hb\ values in the low \ha\ flux distribution are in the x-axis range of log([\nii]/\ha)$\sim$ (-1.3,-1.0). 
These \emph{spaxels} are located close to the stars (north-west of the central clusters), where no high intensity \ha\ emission is observed (see top-right hand panels in Figs~\ref{flux_ha_hb} and ~\ref{lineratio}). 

The data distribution in the [\oiii]/\hb\ versus [\sii]/\ha\ diagram (bottom panels of Fig.~\ref{BPTdiag}) shows a similar but no so strong trend as in the top panel. The separation between the distribution of the high \ha\ flux \emph{spaxels} (bottom-left hand panel in Fig.~\ref{BPTdiag}) and the integrated value for the region, marked in the same plot as a black cross, is remarkable. This shows the effect of the aperture selection in the estimation of the [\sii] fluxes and reflects the differences we have found between the results for the integrated measurements reported here and those given in \citet{Esteban:2009p563}, only covering one of the most intense knots in the region (see Section~\ref{secint}). 

\begin{figure*}
\includegraphics[width=0.7\textwidth]{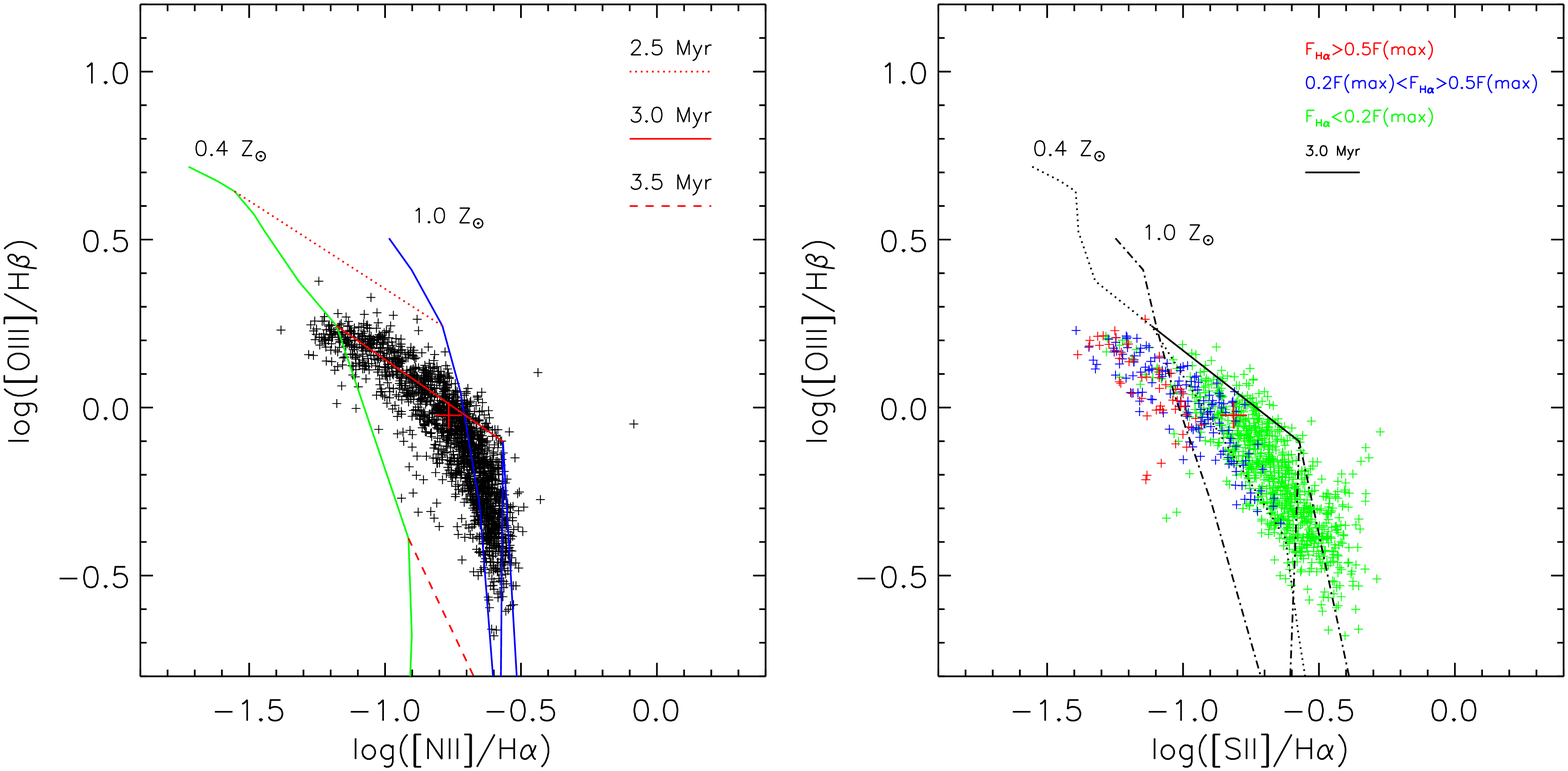}
\caption{BPT emission-line ratio diagnostics as shown in Fig.~\ref{BPTdiag} compared with models from \citet{Dopita:2006p40} (log [(M$_{\rm Cl}$/\msun)/($P_{o}/k$)]=0.0). In the left-hand panel we plot all the \emph{spaxels} without making any distinction in flux, while in the right-hand panel we code the points for different bins of \ha\ flux. We plot the models for two different metallicities 
(Z=\zsun\ and Z=0.4\zsun) covering the range of metallicities reported in Section~\ref{secint}. We also mark the temporal evolution of the region at 2.5, 3.0 and 3.5~Myr.}
\label{BPTmodels}
\end{figure*}

In Fig.~\ref{BPTmodels} we compare the observed BPT diagrams with models from \citet{Dopita:2006p40}, which take into account the effect of the stellar winds on the dynamical evolution of the region. The ionization parameter is replaced by a new variable \emph{R} that depends on the mass of the central cluster and the pressure of the interstellar medium ( \emph{R} = (M$_{\rm Cl}$/\msun)/($P_{o}/k$), with $P_{o}/k$ measured in (\cmtres\ K)). Using an estimate for the mass of the cluster of 2$\times$10$^{5}\msun$ \citep{Relano:2009p558} and the temperature listed in Table~\ref{tabint}, a  density value of 13~\cmtres\ (Section~3) gives log {\it R}=0.3. The models shown in Fig.~\ref{BPTmodels} are for log \emph{R}=0.0. All the \emph{spaxels} fall in the area delimited by the lines at Z=\zsun\ and Z=0.4\zsun, which agree with the estimate of the metallicity of the region from other methods (see Section~\ref{secint}). The age of the region derived from the models ($\sim$3.0-3.5~Myr) also agrees with previous estimates in the literature (4.5$\pm$1.0~Myr \citet{Malumuth:1996p494} and 3.5$\pm$0.5~Myr \citet{Pellerin:2006p499}). 
The location of the points in the right-hand panel in Fig.~\ref{BPTmodels} depends on the \ha\ flux: \emph{spaxels} with high \ha\ flux have generally lower values of [\sii]/\ha, while those with low \ha\ flux are located towards the right in the x-axis. 
 
\subsection{WR stellar population}

WR stars are very bright objects which show strong broad
emission lines in their spectra. They can be classified as nitrogen (WN) stars (those
with strong lines of helium and nitrogen) and carbon (WC) stars (those with strong
lines of helium, carbon and oxygen). They are understood as the result of the
evolution of massive O stars which lose a significant amount of their
mass via stellar winds showing the products of the CNO burning
first (WN stars) and the He burning afterwards (WC stars)
\citep{Conti:1976p480}. The short phase for the WR stars 
makes their detection a very precise method for estimating the age of a
given stellar population. Typically, an instantaneous burst of star
formation shows these features at ages of $\sim2-6$~Myr and even at a
more limited range at very low metallicities \citep{Leitherer:1999p491}.
The presence of WR stars can be recognised via the WR bumps around
$\lambda$4650~\AA{} (i.e. the \emph{blue bump}, characteristic of WN stars) and
$\lambda$5808~\AA{} (i.e. the \emph{red bump},  characteristic of WC
stars).

The WR population in NGC~595 has already been studied in \citet{Drissen:1993p105}
and \citet{Drissen:2008p481} by means of high spatial resolution
narrow imaging of the central area of the nebula, first, and
spectroscopic follow-up, afterwards. They detected 11 WR
candidates: nine of them were spectroscopically classified, while
those named as WR10 and WR11 were removed from the candidate list.
The WR survey done by \citet{Drissen:1993p105} and \citet{Drissen:2008p481} only 
covers the central part of the region with a field of view of 30\arcsec$\times$35\arcsec, and the 
spectral resolution of their observations is $\sim$9~\AA. The observations presented here have 
higher spectral resolution (3.4~\AA) and cover an four times greater. This 
allows us to make a complete census of the WR population of the region. 

\begin{figure}
\includegraphics[width=0.45\textwidth]{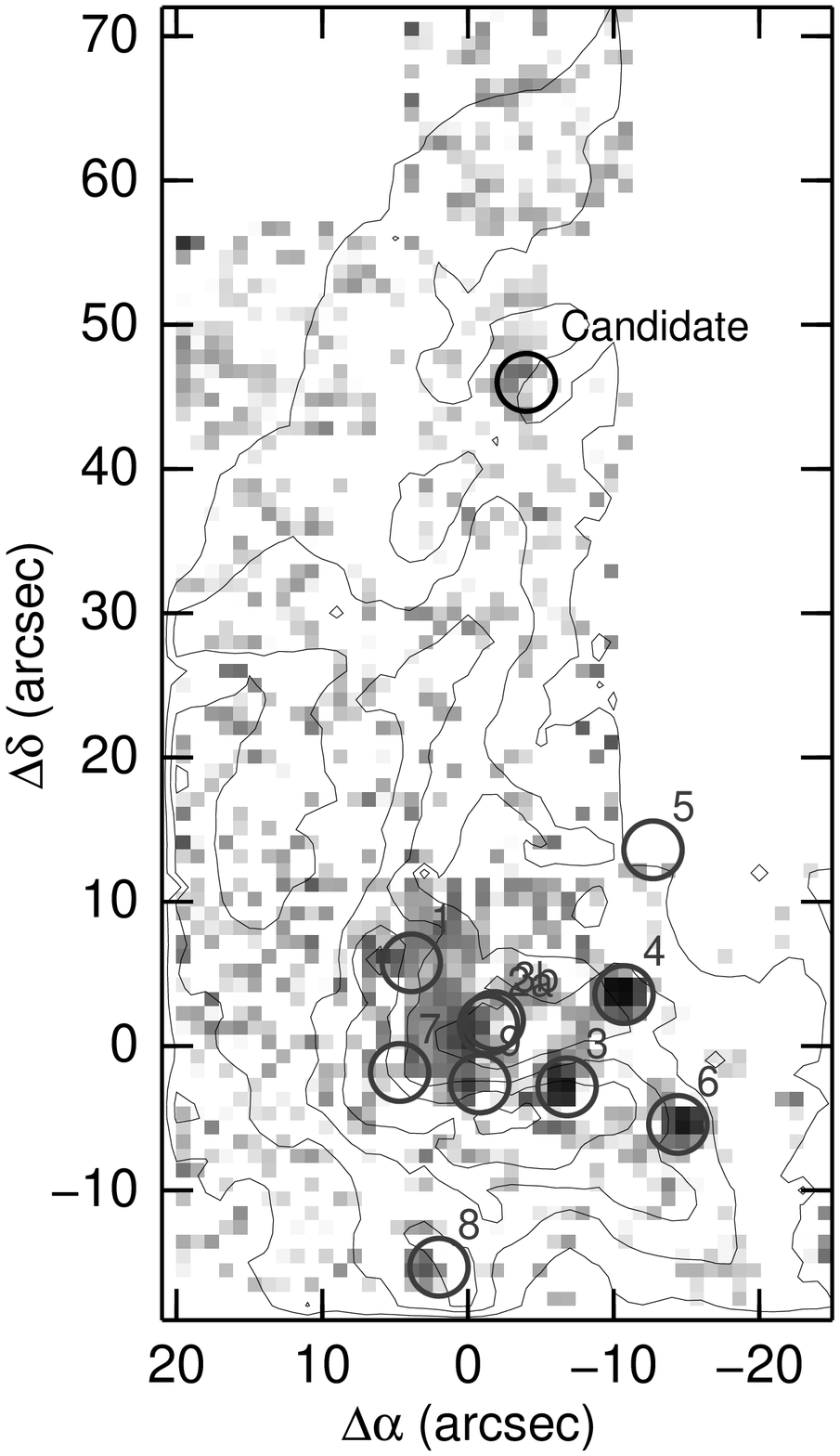}
\caption{4490-4540~\AA{} map derived from the PMAS data with contours of the \ha\ emission (see Fig.~\ref{flux_ha_hb}) 
in logarithmic scale. Continuum has
 been substracted by averaging the spectral ranges of 4650-4750~\AA{}
 and 4755-4805~\AA{}. We have overplotted the positions of the WR
 candidates listed by \citet{Drissen:1993p105} with dark grey circles. 
 Contours are in a logarithmic scale and over an interval of 1.45 dex in steps of 0.36 dex.
 The current data show a new candidate towards the northern part of the nebula
 which has been indicated with a black circle. Orientation is the same as in the rest of the maps and fluxes are in arbitrary units.}
\label{wrposi}
\end{figure}

In order to make a first identification of the WR stars, we constructed continuum-substracted images at
4490-4540~\AA{} and 5645-5695~\AA{} spectral ranges. Fig.~\ref{wrposi} shows the
positions of the WR features detected by \citet{Drissen:1993p105} overplotted on
our continuum-substracted 4490-4540~\AA{} map. There is a good correspondence within the
errors of the astrometry. All the candidates but WR5 (which is not
covered in the current data, as shown in Fig.~\ref{wrposi}) and WR7 (the one with the faintest measured bump) are detected by visual
inspection. In addition, we have detected a new candidate towards the
north of the  \textsc{H\,ii} region. WR3 and WR1 were also detected in the map for the \emph{red bump} (not
shown). This map was less reliable since important residuals in the
5577~\AA{} sky line do not allow to select a reliable spectral range
for the continuum subtraction.

\begin{figure}
\includegraphics[width=0.5\textwidth]{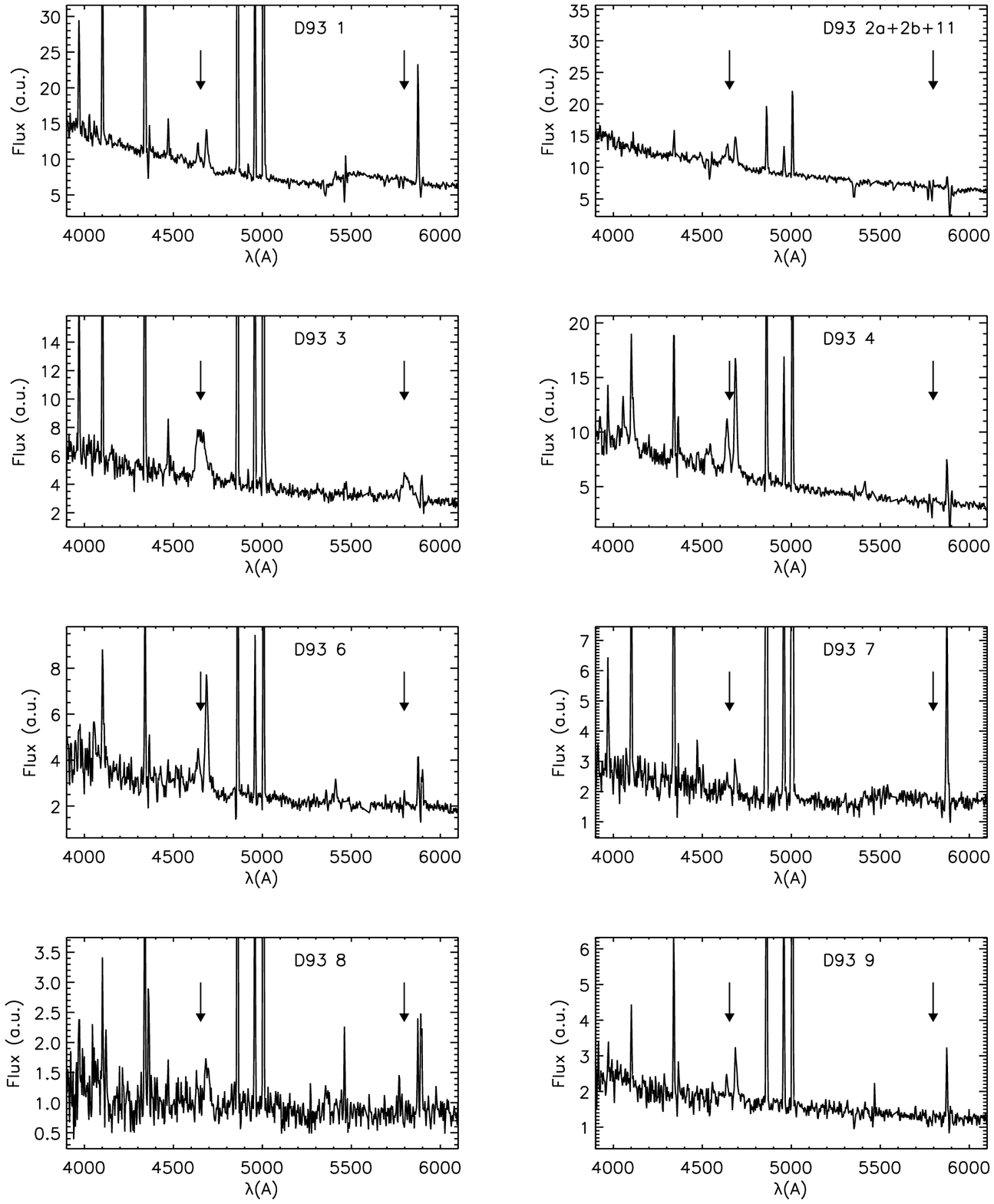}
\includegraphics[width=0.45\textwidth]{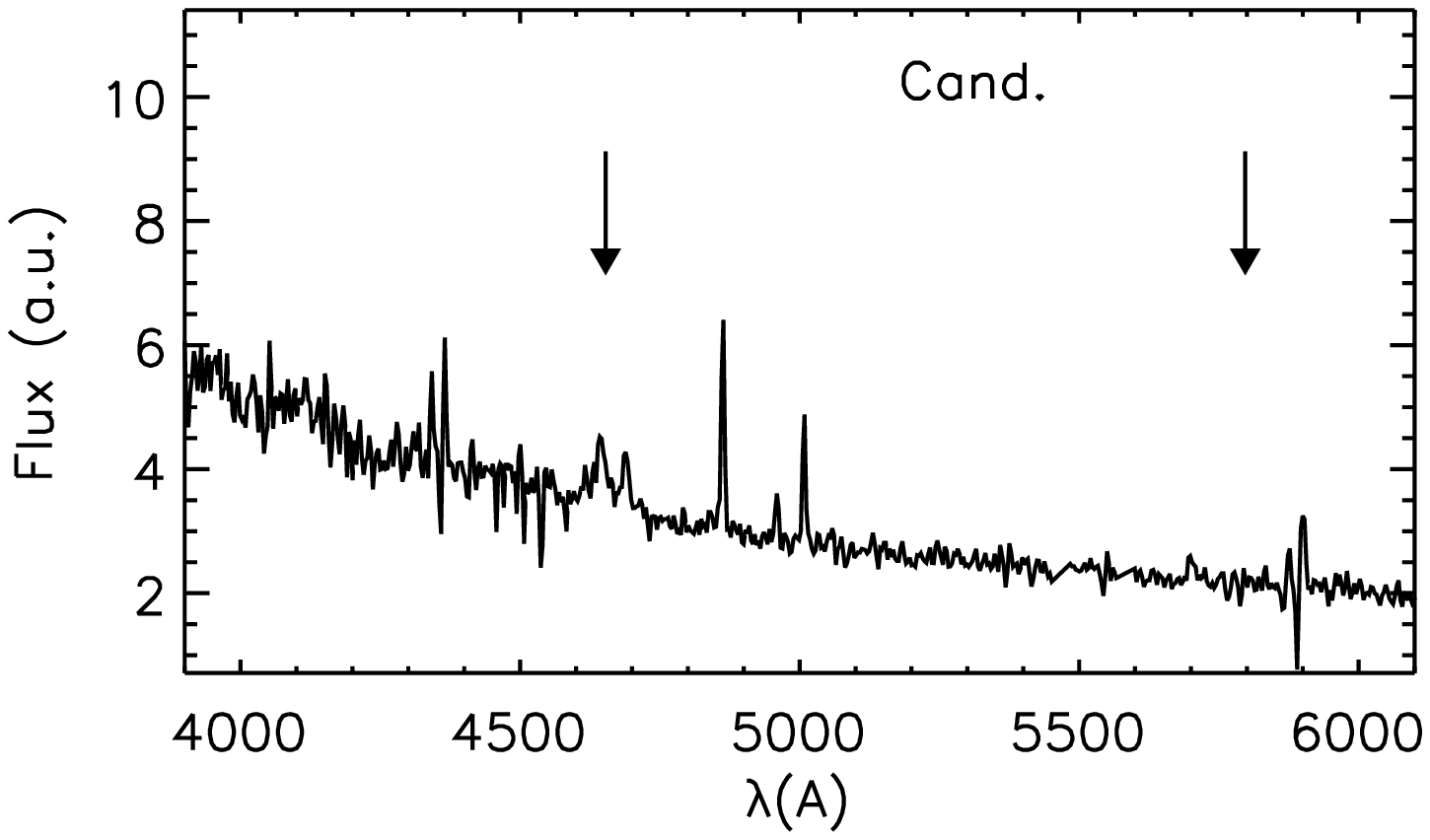}
\caption{Extracted spectrum for the new WR candidate (bottom) and the WR stars catalogued by 
\citet{Drissen:2008p481} (top). Vertical arrows
 mark the position of the expected \emph{blue} and \emph{red bumps}.
 \label{specwr}}
\end{figure}

We extracted the integrated spectra of the WR candidates by coadding typically
$\sim3-10$ spectra centred at the position of the WR detections. Fig.~\ref{specwr} 
shows the spectra for our new candidate (bottom panel), as well as for the 
WR catalogued by \citet{Drissen:2008p481} (top panel), and illustrates the power of the IFS data to identify and characterize WR stars. In all the spectra shown in Fig.~\ref{specwr} 
the \emph{blue bump} characteristic of WN populations is clearly visible. We compare our spectra with those shown in \citet{Drissen:2008p481}. These authors present the spectra of four WR stars, except for WR5 (not covered by our observations) the spectra of the other stars (WR1, WR3 and WR9) show the same features as those 
presented by \citet{Drissen:2008p481}. We confirm the classification of the WR stars catalogued by these authors, all of them being WN stars except WR3, whose red bump is clearly observed at 5800~\AA, showing that it is a WC star.

We estimate the EW for the \emph{blue bump} as a sum of the individual EW for the emission lines seen in the 4600-4700~\AA\ wavelength range. 
Except for WR3 and WR4, the EW(\emph{blue bump}) of the stars in Fig.~\ref{specwr} ranges within 15-40~\AA, which agrees with the models of \citet{Schaerer:1998p553} (see their fig. 11 for metallicities of Z=0.08 and Z=0.02). 
We derive an EW(\emph{blue bump})=28~\AA\ for WR9, similar to the value reported by \citet{Drissen:2008p481}. For WR3 and WR4 we find EW(\emph{blue bump})$\sim$60~\AA, which are consistent with previous values given in the literature (\citet{Massey:1983p556} and \citet{Armandroff:1991p557}, respectively). 

The spectrum of our WR candidate is shown in the bottom panel of Fig.~\ref{specwr}. The \emph{blue bump} is clearly seen in the spectra but no red bump is visible, which shows that this star is probably a WN star. We derive an EW(\emph{blue bump})=14~\AA\ for our new candidate, which is within the expected range of values given in \citet{Schaerer:1998p553}. We cannot rule out the possibility of this star being an Of rather than a WR star (P. Crowther, private communication), but further spectroscopic observations at higher resolution are required to assess this issue. It is interesting to note the isolation of our new WR candidate at a distance of $\sim$185~pc away from the location of the rest of the WR stars. Assuming a time of 3-4~Myr, consistent with the age of the \hii\ region, we find radial velocities of 45-60~\kms \ for the WR candidate. These velocities are in the range of values expected for runaway WR and O stars (\citealt{Moffat:1998p566}; \citealt{Dray:2005p567}; \citealt*{Hoogerwerf:2001p572}).

\subsection{Metallicity, ionization parameter and structure of the region}
\label{metstruc}
It is well known that the \rdostres\ index (\rdostres = ([\oii] $\lambda$3727 +  [\oiii] $\lambda\lambda$4959,5007)/\hb) does depend on the ionization parameter but the question that arises is how strong is the variation of this parameter within NGC~595? Several observational studies have claimed evidence for a relative constant value of \rdostres\ in spite of the strong variations of the ionization parameter within \hii\ regions (\citealt{Oey:2000p121}, \citealt{Oey:2000p124}; \citealt{Kennicutt:2000p110}). Recently, \citet{Ercolano:2007p482} suggest that the geometrical distribution of the ionizing stars could account for the discrepancies in the metallicities derived using classical metallicity calibrators. In order to shed light on this problem, we have studied the variations across the surface of NGC~595 of several classical 
emission-line ratios proposed as metallicity calibrators and compared them with the variation of [\oiii] $\lambda\lambda$4959, 5007/[\oii] $\lambda$3727, a tracer of the ionization parameter.  

\begin{figure*}
\vspace{-1.5cm}
\includegraphics[width=0.4\textwidth]{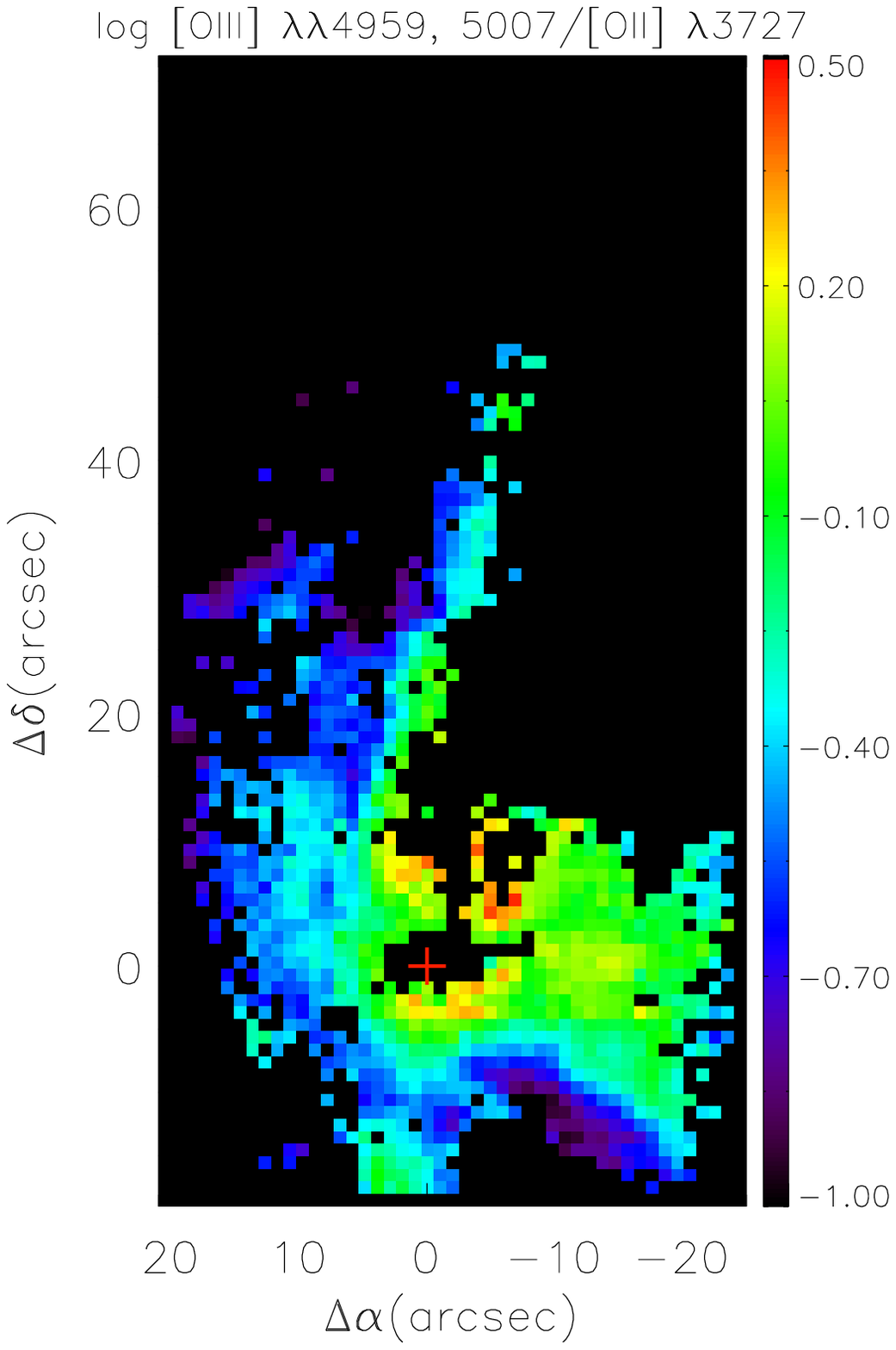}
\hspace{1cm}
\includegraphics[width=0.4\textwidth]{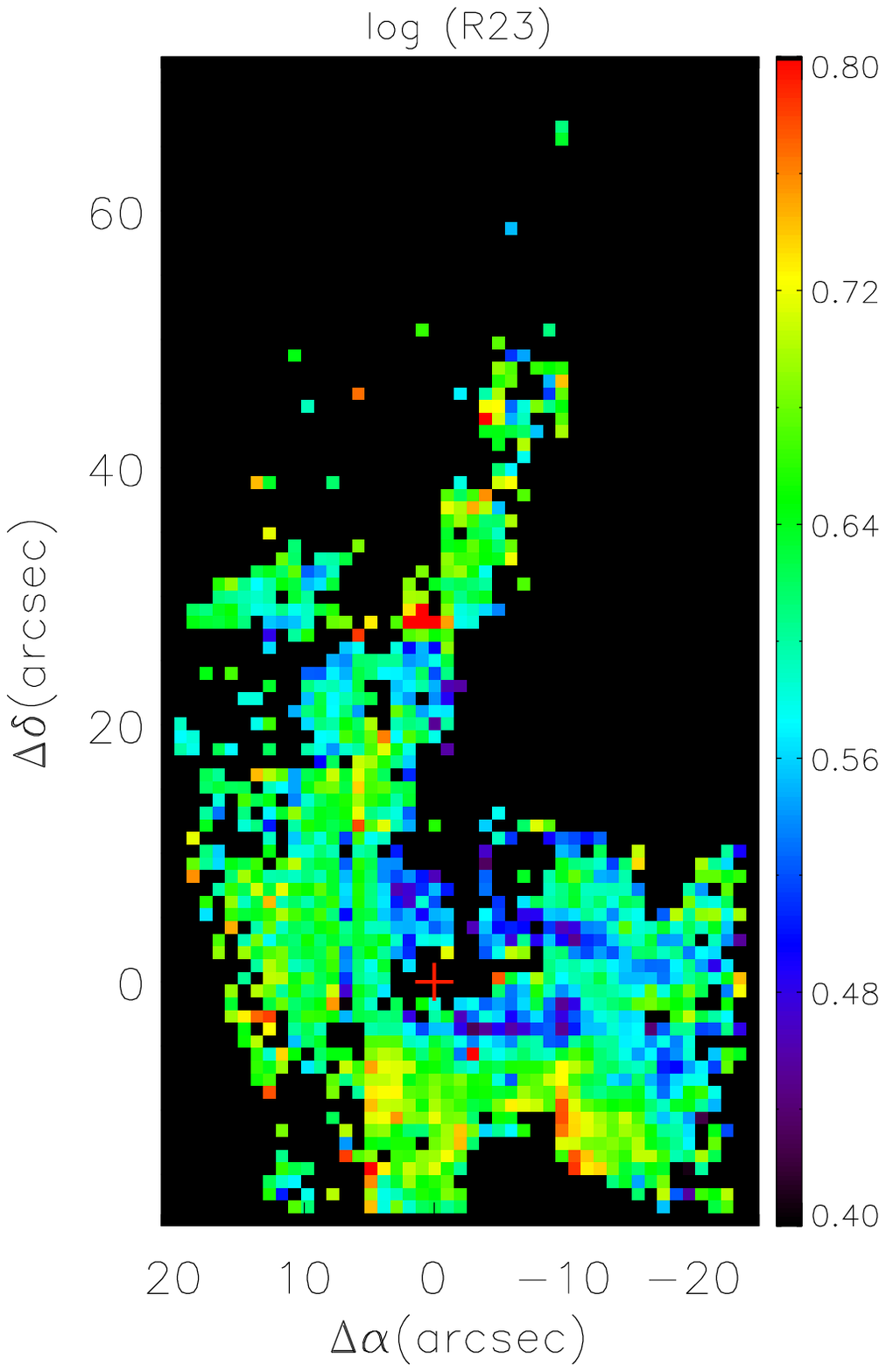}
\includegraphics[width=\textwidth]{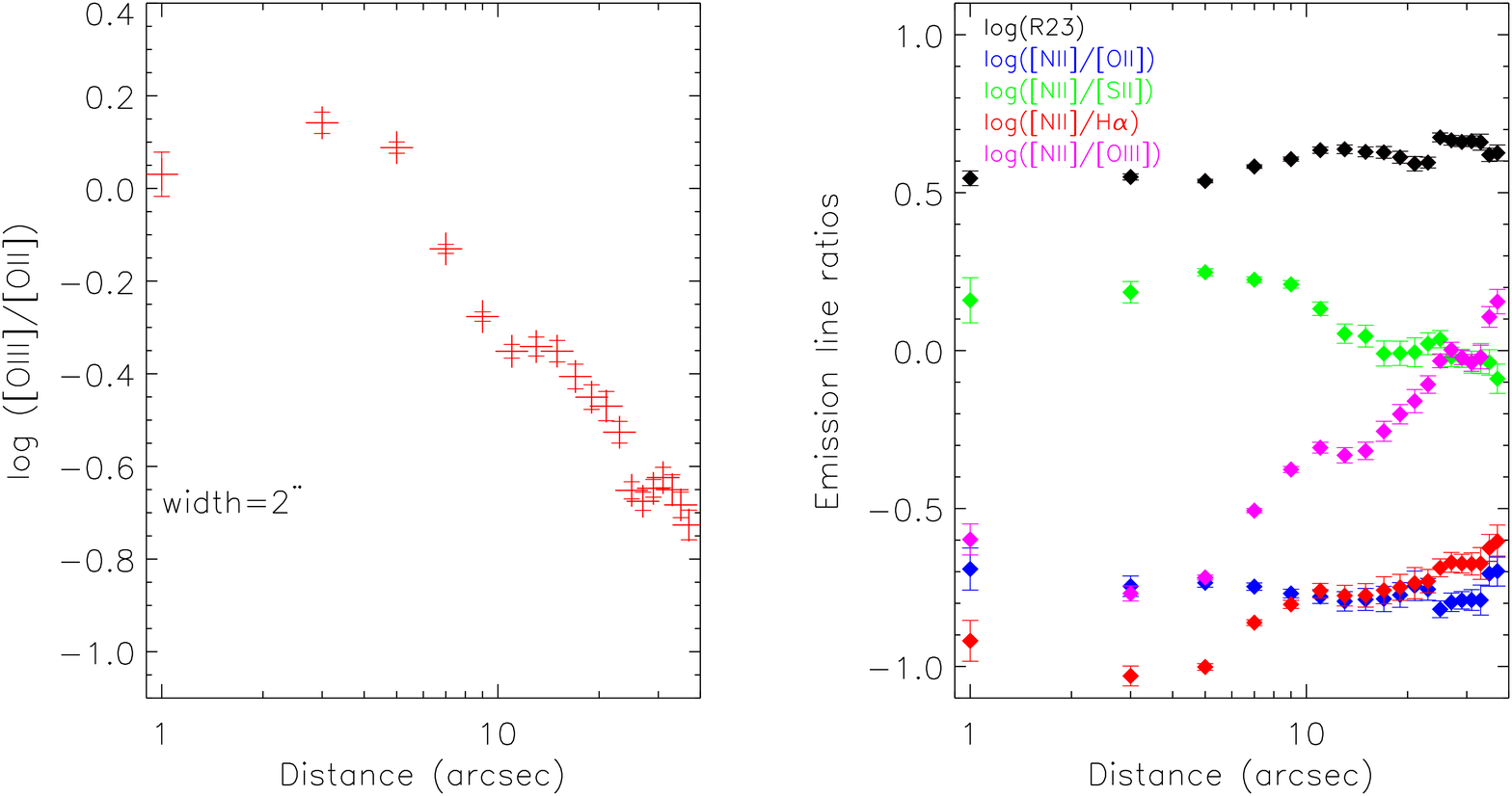}
\caption{Top: maps of the oxygen emission-line ratios ([\oiii] $\lambda\lambda$4959, 5007/[\oii] $\lambda$3727) tracing the ionization parameter (left-hand panel) 
and the metallicity (\rdostres = ([\oii] $\lambda$3727 +  [\oiii] $\lambda\lambda$4959,5007)/\hb) (right-hand panel). The emission-line ratios shown here are those 
having relative errors $<$ 30 per cent. The orientation is the same as in Fig.~\ref{lineratio}. Bottom: elliptical radial profiles of [\oiii]/[\oii] (left-hand panel) and \rdostres\ (right-hand panel, black diamonds). The ellipse parameters are the same as the one used in Figs.~\ref{perfextratio}  and ~\ref{perfsiiratio} and the rings have 2 arcsec width. In the right-hand panel, we also plot radial profiles of the main emission-line ratios used as metallicity calibrators: [\nii]/[\oii] (blue), [\nii]/[\sii] (green), [\nii]/\ha\ (red) and [\nii]/[\oiii] (magenta).}
\label{ratox_r23}
\end{figure*}

In the top panels of Fig.~\ref{ratox_r23}, we show maps of the [\oiii] $\lambda\lambda$4959, 5007/[\oii] $\lambda$3727 (left) and \rdostres\ (right). There is a clear radial trend of [\oiii] $\lambda\lambda$4959, 5007/[\oii] $\lambda$3727 with the radial distance from the central stars: close to the ionizing stars we find high values of this ratio, implying a high ionization parameter, while moving towards larger distances from the ionizing stars the emission-line ratio declines, showing that the ionization parameter is lower at these locations. 
A map of the \rdostres\ index for the whole region is shown in Fig.~\ref{ratox_r23} (top right). We have quantified the variations for both maps in the lower panels of Fig.~\ref{ratox_r23}. Using the same elliptical integration as we have explained in Section~\ref{extinc_section}, we derived the values for [\oiii] $\lambda\lambda$4959, 5007/[\oii] $\lambda$3727 and \rdostres\ in elliptical annuli of 2 arcsec width. 
While there is a strong gradient of 1 order of magnitude from the centre to the outer parts of the region for [\oiii] $\lambda\lambda$4959, 5007/[\oii] $\lambda$3727 (Fig.~\ref{ratox_r23} bottom left), \rdostres\ (black diamonds in the bottom right-hand panel) shows a variation of 0.13~dex over the whole radial distance. This shows the reliability in using \rdostres\ as an emission-line ratio to obtain the metallicity of this \hii\ region; even when only a part of the region is covered by the observations, which is normally the case for long-slit spectroscopy, the value derived for \rdostres\ can be used as a representative one for the \hii\ region. The variation of  \rdostres\ in Fig.~\ref{ratox_r23} (bottom right) would translate into a nominal difference of 0.17~dex in log[O/H] in the case we would use the calibration of \citet{McGaugh:1991p514}.

For comparison, we have included in Fig.~\ref{ratox_r23} (bottom right) other emission-line ratios proposed in the literature as metallicity calibrators: [\nii] $\lambda$6584/[\oii] $\lambda$3727 (blue diamonds), [\nii] $\lambda$6584/[\sii] $\lambda\lambda$6717, 6731 (green diamonds),  [\nii] $\lambda$6584/\ha\ (red diamonds), and [\nii] $\lambda$6584/[\oiii] $\lambda$5007 (magenta diamonds). The figure shows that  [\nii]/\ha\ and [\nii]/[\oiii] vary significantly with the radial distance. The variation is particularly significant for [\nii]/[\oiii], 0.92~dex, which is expected since this ratio strongly depends on the ionization parameter \citep{Kewley:2002p405}. The variations of [\nii]/\ha\ and [\nii]/[\oiii] presented here show a similar trend as those given by \citet{Ercolano:2007p482} for different configuration of stars and gas in \hii\ regions at the metallicity of NGC~595 (Z= (0.4-1)~\zsun\ in the models of \citealt{Ercolano:2007p482}). Following the calibrations of \citet{Pettini:2004p551} for [\nii]/\ha\ and [\nii]/[\oiii] we find variations of $\sim$0.3 in log[O/H]  for both emission-line ratios. Finally, [\nii]/[\sii] and [\nii]/[\oii] are more stable within the region; in particular [\nii]/[\oii] would show a rather similar behaviour as \rdostres, which would make it as good emission-line ratio to trace the metallicity of the regions as \rdostres\ is.

\section{Conclusions}

We present IFS of NGC~595, one of the most luminous \hii\ regions in M33, covering an unprecedented area in the disc of the spiral galaxies of the Local Group, a $\sim$174$\times$340~parsec$^2$ field of view representing the complete surface of the region. Taking advantage of the power of these observations, we are able to identify and catalogue the WR population of the region and to make the best census of WR stars in NGC 595 up till now. We have analysed the variations within the region of the main emission-line ratios that describe the physical properties of the region. The analysis of the observations yields the following results.

\begin{itemize}

\item We present the properties of NGC~595 derived from the integrated spectrum of the region. The electron density and 
the metallicity are consistent with previously reported values in the literature using long-slit spectroscopy, covering just the most intense knots in the region. 

\item The extinction map obtained from the \ha/\hb\ emission-line ratio at each \emph{spaxel} in the field of view presents a concentrated distribution with the maximum located at the centre of the \ha\ shell structure. We have compared this map with the {\it Spitzer} 24 and 8~\mi\  bands using elliptical radial profiles. The 24~\mi\ emission and the \ha/\hb\ radial profiles are similar, with their maxima located at the same distance from the ionizing stars. The maximum of the 8~\mi\ emission radial profile shows a displacement of $\sim$4-5\arcsec\ (15-18~pc) with respect to the maximum of the \ha/\hb\ radial profile. This shows that the extinction suffered by the gas is produced by dust emitting at 24~\mi, which is probably mixed with the ionized gas within the region, while the dust emitting at 8~\mi\ is more related to the outer parts of the region where the molecular cloud is located. 
 
\item The [\sii]$\lambda$6717/[\sii]$\lambda$6731 emission-line ratio map does not show any structure, implying that the electron density is quite constant within the region, despite the pronounced \ha\ shell morphology of NGC~595. The values of [\sii]$\lambda$6717/[\sii]$\lambda$6731 are consistent with the low electron density limit. 

\item We have produced BPT diagrams using the emission-line ratios at each \emph{spaxel} in the field of view. We find that the \emph{spaxels} with a low \ha\ flux cover the total excitation range of the region. The location of the \emph{spaxels} with a high \ha\ flux in the [\oiii]/\hb-[\sii]/\ha\ diagram is displaced with respect to the position given in the diagram for the integrated values. This shows the limitations of the long-slit observations, which usually cover the most intense \ha\ knot in the regions, to obtain a reliable value for the [\sii]/\ha\ emission-line ratio.  

\item The study of the WR population for NGC~Ê595 has been completed with these observations. We have covered the total surface of the region and found a new WR candidate whose spectrum is characteristic of a WN star. The WR candidate, situated far away from the other WR stars in the region, is close to the \ha\ shell but corresponds to a zone of low \ha\ emission. 

\item  We have analysed the behavior of the most used metallicity calibrators: \rdostres\ shows small variations of 0.13~dex despite the spatial distribution of the stars and gas in the region and the strong trend of [\oiii]/[\oii] to decrease to the outer parts of the region. This shows the robustness of \rdostres\ to estimate the metallicity of the region; in spite of the \hii\ region morphology, far away from the classical Str\"omgren sphere picture, the \rdostres\ parameter varies slightly within the region and thus can be used to obtain a representative value of the \hii\ region metallicity. In contrast, other parameters such as [\nii]/\ha\ and [\nii]/[\oiii] show strong variations within the region (up to one order of magnitude in the case of [\nii]/[\oiii]), which show their strong dependence on the ionization parameter.
\end{itemize}

We show in this paper the power of the IFS, which overcomes the limitations of long-slit observations in \hii\ regions. The results presented here show that for the emission lines dominated by the contribution of the most intense knots, long-slit observations are representative of the integrated emission of \hii\ regions. 
The situation is different in the case of lower excitation emission lines, as it happens in the [\sii] emission lines, for which there is a significant bias for long-slit observations. This bias has only been possible to quantify using IFS observations with a complete \hii\ region coverage. 

The physical properties can vary within \hii\ regions and the emission-line ratio maps derived from IFS observations allow us to make a detailed study of these variations. For NGC~595 we find that the ionization parameter shows strong variations with the radial distance from the stars; the ionization structure is clearly depicted with the corresponding emission-line diagnostics and the reddening map presents a non-uniform distribution with a maximum located at the center of the observed \ha\ shell structure. The electron density map, however, does not present significant variations within the region. 

A change of one order of magnitude in [\oii]/[\oiii] within NGC~595 allows us to study the dependence of the ionization parameter on the most widely used metallicity calibrators, since other properties such as density or the ionizing spectrum do not change within this \hii\ region. The main result is that \rdostres\ does not depend significantly on the ionization parameter, but other emission-line ratios such as [\nii]/\ha\ and [\nii]/[\oiii] show important variations within the region.

Finally, the capability of the IFS data to make complete censuses of the WR population in stellar clusters is very nicely shown here: while the hunting of WR stars are based on identification in broad-band images with follow-up spectroscopic observations, we are able to perform such an analysis in only one night of observation, covering the whole surface of the region and allowing us to identify WR stars that are not located close to the centre of the cluster.

\section{Acknowledgments}
We thank the referee for helpful comments to improve the new version of this paper. The authors would like to thank John Eldridge, Paul Crowther and Enrique P\'erez-Montero for constructive comments and useful discussions. 
This research was supported by a Marie Curie Intra European Fellowship within the 7$^{\rm th}$ European Community Framework Programme. 
AMI is supported by the Spanish Ministry of Science and Innovation (MICINN) under the program "Specialization in
International Organisms", ref. ES2006-0003. JMV acknowledges partial funding through research projects
   AYA2007-67965-C03-02 from the Spanish PNAYA  and CSD2006 00070 ¬1st Science with GTC¬ of the MICINN. This research draws upon data provided by {\it The Resolved Stellar Content of Local Group Galaxies Currently Forming Stars} PI: Dr. Philip Massey, as distributed by the NOAO Science Archive. NOAO is operated by the Association of Universities for Research in Astronomy (AURA), Inc. under a cooperative agreement with the National Science Foundation. Some of the data presented in this paper were obtained from the Multimission Archive at the Space Telescope Science Institute (MAST). STScI is operated by the Association of Universities for Research in Astronomy, Inc., under NASA contract NAS5-26555. Support for MAST for non-HST data is provided by the NASA Office of Space Science via grant NAG5-7584 and by other grants and contracts. This paper uses the plotting package jmaplot, developed by Jes\'us Ma\'{\i}õz-Apell\'aniz. http:dae45.iaa.csic.es:8080$\sim$jmaiz/software.

\bibliographystyle{mn2e}
\bibliography{mnras} 

\label{lastpage}

\end{document}